\documentclass[journal,draftcls,onecolumn,12pt,twoside]{IEEEtran}
\usepackage{amsmath,amssymb,threeparttable,placeins,makecell,stfloats,cite}
\usepackage[dvips]{graphicx}
\usepackage[usenames]{color}
\usepackage{amsfonts}
\usepackage{latexsym}
\usepackage{subfigure}
\usepackage[justification=centering]{caption}
\usepackage{floatrow}
\newtheorem{theorem}{{{\textit{Theorem}}}}

\newtheorem{lemma}{{{\textit{Lemma}}}}
\newtheorem{corollary}{{{{\textit{Corollary}}}}}

\newtheorem{definition}{{{\textit{Definition}}}}

\newtheorem{remark}{{{\textit{Remark}}}}

\newtheorem{example}{{{\textit{Example}}}}

\begin{document}
\title{Optimal Z-complementary Code Set From Generalized Reed-Muller Codes}
\author{Palash~Sarkar,~
        Sudhan~Majhi,~
        and~Zilong~Liu
\thanks{Palash Sarkar is with Department of Mathematics and Sudhan Majhi is with the Department of Electrical Engineering, Indian Institute of Technology Patna, India, e-mail: {\tt palash.pma15@iitp.ac.in; smajhi@iitp.ac.in}.}
\thanks{Zilong Liu is with Institute for Communication Systems, 5G Innovation Centre, University of Surrey, UK, e-mail: {\tt zilong.liu@surrey.ac.uk}.}}

\IEEEpeerreviewmaketitle
\maketitle
\begin{abstract}
Z-complementary code set (ZCCS), an extension of perfect complementary codes (CCs), refers to a set of two-dimensional matrices having zero
correlation zone properties. ZCCS can be used in various multi-channel systems to support, for example, quasi-synchronous interference-free multicarrier
code-division multiple access communication and optimal channel estimation in multiple-input multiple-output systems. Traditional
 constructions of ZCCS heavily rely on a series of sequence operations which may not be feasible for rapid hardware generation particularly
 for long ZCCSs. In this paper, we propose a direct construction of ZCCS using second-order Reed-Muller codes with efficient graphical representation. Our proposed construction, valid for any number of isolated vertices present in the graph, is capable of generating optimal ZCCS meeting the set size upper bound.
\end{abstract}

\section{Introduction}\label{sec:intro}
\IEEEPARstart{C}{ode}-division multiple-access (CDMA) technology is an important multiuser communication scheme where spreading
sequences play a fundamental role in determining the system performance. Traditional  spreading sequences, such as Walsh-Hadamard
sequences, pseudo-random sequences (e.g., Gold sequences, Kasami sequences, optimal $\mathbb{Z}_4$ sequences), constant amplitude zero
auto-correlation (CAZAC) sequences, generally exhibit nonzero cross-correlation properties over asynchronous
transmission channels. Because of this, their corresponding CDMA systems may suffer from severe ``near-far effect"
whereby the desired signals could be overwhelmed by multiple-access interference (MAI). In legacy CDMA systems (e.g., 3G), tedious power
control is applied to suppress the near-far effect. In this paper, we are focused on Z-complementary code set (ZCCS) which
is capable of supporting interference-free multicarrier CDMA (MC-CDMA) in quasi-synchronous channels (without the need of power control) \cite{chen2007next}.

In \cite{gol1961}, M. J. E. Golay proposed a pair of sequences, known as Golay complementary pair (GCP), with the property that the sum
of their aperiodic auto-correlation function (AACF) is zero everywhere except at the zero-shift position. Either sequence in a GCP is called a Golay sequence. In \cite{chong1972}, Tseng and Liu extended the idea of GCP to complementary code (CC) each consisting of two or more constituent sequences with the same AACF property. Davis and Jedwab proposed a direct construction of GCP from
generalized Boolean function (GBF) to reduce the peak-to-mean envelope power ratio (PMEPR) of orthogonal frequency division multiplexing (OFDM)
system \cite{Davis1999,Li2010,Liu2013}. As a generalization of the Davis-Jedwab construction, Paterson proposed a construction of CC in \cite{pater2000} by associating each CC
with a graph\footnote{Although Paterson's construction is limited to second-order generalized Reed-Muller (RM) codes, generalization to higher-order ones can be found in \cite{kusch}.}. In \cite[Th. 24]{pater2000}, it is found that after applying deletion operation to several vertices of certain graphs, if the
resulting graph consists of a path and one isolated vertex, then the code corresponding to the graph is a CC. Paterson's idea was further extended
by Rathinakumar and Chaturvedi \cite{arthina} for mutually orthogonal Golay complementary
sets (MOGCS) which are also called complete complementary codes (CCCs) in this paper. Formally, CCC refers to a
collection of CCs where each CC is a two-dimensional matrix  (called a complementary matrix) and any two distinct CCs have
zero aperiodic cross-correlation sums. The Rathinakumar-Chaturvedi construction, however, gives little information on the code
generation when some isolated vertices are present (after deletion operation) in the associated graph. Recently, a new class of
CCC has been introduced in \cite{uda2014} for multi-carrier
code division multiple access (MC-CDMA) with column sequence PMEPR of at most $2$. This is achieved by properly
designing CCs (complementary matrices) such that every column sequence of a complementary matrix is a Golay sequence. The application
of CCC has been extended to interference-free MC-CDMA communication by designing a fractional-delay resilient receiver \cite{Liu_FDRR_2015}.

A drawback of CCC is that the set size is limited by the number of row sequences in each complementary matrix \cite{uda2014,Liu_FDRR_2015,cccsmajhi,slett}.
This problem can be fixed by ZCCS whose aperiodic auto- and cross-correlation functions display zero correlation
zone (ZCZ) properties and whose set size is several times of that of CCC \cite{liu2011}. The ZCZ properties of ZCCS allow
MAI mitigation provided that all the received multiuser signals are roughly synchronous within the ZCZ width \cite{Liu-ITW2014}. In the
literature, binary ZCCSs were first
introduced by Fan \emph{et~al. } \cite{fan2007} and later were extended to generalized pairwise ZCCSs by
Feng \emph{et~al.} \cite{lfengispl2008} for power efficient quadrature carrier modems. There are another type set of codes, introduced in \cite{jli_igc_2008,sarkar_igc}, known as
inter-group complementary code set which can be derived as special case of ZCCS.
In addition to their applications in
MC-CDMA \cite{jli_igc_2008}, ZCCS have also been employed as optimal training sequences in multiple-input multiple-output (MIMO)
communications \cite{fan2008,hmwang2007}.

In this paper, we propose a direct construction of optimal  ZCCS
from second-order cosets of the $q$-ary generalization of the first-order RM codes through a
graphical representation. Specifically, we first  construct a set of $2^{k+p}$ codes, each containing $2^{k+1}$ constituent sequences of length $2^m$. These codes are characterized by a
graph (consisting of $m$ vertices in total) with the property that
deleting $k$ vertices and their associated edges, the entire graph reduces to  a path over $m-k-p$ vertices (where all the relevant edges have
identical weight of $q/2$) and $p$ isolated
vertices. Then, we construct another set of codes by reversing
and taking conjugate of the first set of codes. It is interesting to note that the cross-correlation  between any two codes from
different sets is zero everywhere and the union of these two sets gives a ZCCS of size $2^{k+p+1}$. Our proposed construction is flexible
in that the ZCZ width and set size
of the proposed ZCCS can be varied by freely changing the number of isolated vertices (i.e., $p$) in the graph. It is shown that the CCC in \cite{arthina} is a special case of our proposed construction when the number of isolated vertices is set to zero, i.e., $p=0$.
It is noted that our proposed construction generates ZCCS directly based on GBF and does not rely on any recursive sequence operations. Hence, the proposed construction
is suitable for rapid hardware generation particularly for long ZCCSs. An efficient hardware generator (based on logic AND gates, selectors, and adders)
for 16-QAM almost-complementary sequences can be found in \cite[Example 1]{16qamliu}. We also point out that the CCC in \cite{uda2014}, similar
to \cite{arthina}, are characterized by a graph which after deleting some vertices and their associated edges, constitutes one path but with no
isolated vertex. This is a major difference with our proposed construction for ZCCS in graph representation.

The remainder of the paper is organized as follows. In Section II, some useful notations and definitions are given. In Section
III, a construction of ZCCS is presented and its optimal condition is derived. Later, the proposed ZCCS construction is illustrated by an example. Finally, concluding remarks are drawn in Section IV.
\section{Preliminary}
\label{sec:back}
\subsection{Definitions of Correlations and Sequences}
Let $\textbf{a}=(a_0,a_1,\cdots, a_{L-1})$ and $\textbf{b}=(b_0,b_1,\cdots, b_{L-1})$ be two complex-valued sequences of equal length $L$. For an integer $\tau$, define
\begin{equation}\label{equ:cross}
C(\textbf{a}, \textbf{b})(\tau)=\begin{cases}
\sum_{i=0}^{L-1-\tau}a_{i+\tau}b^{*}_i, & 0 \leq \tau < L, \\
\sum_{l=0}^{L+\tau -1} a_ib^{*}_{i-\tau}, & -L< \tau < 0,  \\
0, & \text{otherwise},
\end{cases}
\end{equation}
and $A(\textbf{b})(\tau)=C(\textbf{b},\textbf{b})(\tau)$.
These functions are called aperiodic CCF (ACCF)
between $\textbf{a}$ and $\textbf{b}$ and the AACF of $\textbf{b}$, respectively.
Let $\textbf{C}=\{C_0,C_1, \cdots ,C_{K-1}\}$ be a set of $K$ matrices (codes), each having order $M\times L$ as follows.
\begin{equation}
\begin{split}
C_\mu=\begin{bmatrix}
\textbf{a}_0^\mu \\ \textbf{a}_1^\mu \\ \vdots \\ \textbf{a}_{M-1}^{\mu}
\end{bmatrix}_{M\times L},
\end{split}
\end{equation}
where $\textbf{a}_\nu^\mu$ $(0\leq \nu \leq M-1,0 \leq \mu \leq K-1)$ is the $\nu$-th row sequence or $\nu$-th constituent sequence  of $C_\mu$.
Let $C_{\mu_1}$, $C_{\mu_2}\in \textbf{C}$ $(0\leq \mu_1,\mu_2\leq K-1)$ be any two matrices in $\textbf{C}$. The ACCF
of $C_{\mu_1}$ and $C_{\mu_2}$ is defined by
 \begin{equation}
  C(C_{\mu_1},C_{\mu_2})(\tau)=\displaystyle \sum_{\nu=0}^{M-1}C(\textbf{a}_\nu^{\mu_1},\textbf{a}_\nu^{\mu_2})(\tau).
 \end{equation}
\begin{definition}
 Code set $\textbf{C}$ is said to be a set of  CCC if $K=M$ and
 \begin{equation}
  C(C_{\mu_1},C_{\mu_2})(\tau)
=\begin{cases}
LK, & \tau=0, \mu_1=\mu_2;\\
0, & 0<|\tau|<L, \mu_1=\mu_2;\\
0, & |\tau|< L, \mu_1\neq \mu_2.
\end{cases}
 \end{equation}
\end{definition}
In the above definition, each code $C_\mu$ $(0\leq \mu \leq K-1)$, is said to be a CC. When $M=2$, $C_\mu$
reduces to a GCP and either sequence of the pair is called a Golay sequence.
\begin{definition}
Code set $\textbf{C}$ is called a ZCCS denoted by $(K,Z)$-$\text{ZCCS}_M^L$ if
\begin{eqnarray}
C(C_{\mu_1},C_{\mu_2})(\tau)
=\begin{cases}
LM, & \tau=0, \mu_1=\mu_2,\\
0, & 0<|\tau|<Z, \mu_1=\mu_2,\\
0, & |\tau|< Z, \mu_1\neq \mu_2,
\end{cases}
\end{eqnarray}
 where $Z$ is called ZCZ width.
 \end{definition}
\subsection{Generalized Boolean Functions}
Let $f$ be a function of $m$ variables $x_0,x_1,\cdots,x_{m-1}$ over $\mathbb{Z}_q$. A monomial of degree $k$ is defined as the product
of any $k$ distinct variables among  $x_0,x_1\cdots x_{m-1}$. There are $2^m$ distinct monomials over $m$ variables listed below:
\begin{equation}
\begin{split}
1,x_0,x_1,\cdots,x_{m-1},x_0x_1,x_0x_2,\cdots,x_{m-2}x_{m-1},\cdots,\\x_0x_1\cdots x_{m-1}.
\end{split}
\end{equation}
A function $f$ is said to be a GBF if it can uniquely be expressed as a linear
combination of these $2^m$ monomials, where the coefficient of each monomial is drawn from $\mathbb{Z}_q$.  Corresponding to
each GBF $f$, we define a complex valued sequence $\psi(f)$ of length $2^m$ by defining
$\psi(f)=(\omega^{f_0}, \omega^{f_1}, \cdots, \omega^{f_{2^m-1}})$, where $f_i=f(i_0,i_1,\cdots,i_{m-1})$, $\omega=\exp(2\pi\sqrt{-1}/q)$ $(q$ is a positive integer no less than  $2)$
and $(i_0,i_1,\cdots,i_{m-1})$ is the binary vector representation of integer $i$ $(i=\sum_{j=0}^{m-1}i_j2^j)$. We denote by $\bar{x}=1-x$
the binary complement of $x\in \{0,1\}$.
For any given GBF $f$ in $m$ variables, we denote the function $f(1-x_0,1-x_1,\cdots,1-x_{m-1})$ or $f(\bar{x}_0,\bar{x}_1,\cdots,\bar{x}_{m-1})$  by $\tilde{f}$.
For a complex-valued sequence $\textbf{a}$, let $\tilde{\textbf{a}}$ denote the sequence obtained by reversing $\textbf{a}$ and $\textbf{a}^*$ its complex conjugate.
\subsection{Some Families of Codes}
A linear code over $\mathbb{Z}_q$ of length $L$ is closed
under linear combinations of sequences (called codewords). Corresponding to any such code $\zeta$ there is a generator matrix
$\text{G}$. Linear combinations of the rows of $\text{G}$ generate the code. For any fixed sequence $\textbf{a}$ of length $L$, $\textbf{a}+\zeta$
denotes a coset of $\zeta$ and $\textbf{a}$ is said to be a coset representative of $\zeta$. RM$_q(r,m)$ is said to be the $r$th order RM code
whose codewords are $\mathbb{Z}_q$-valued sequences identified with  GBFs of degree
at most $r$ in $x_0, x_1, \cdots, x_{m-1}$. The rows of generator matrix $\text{G}$ for RM$_q(r,m)$ are $\mathbb{Z}_q$-valued sequences corresponding
to distinct monomials of degree at most $r$ over the variables $x_0, x_1, \cdots, x_{m-1}$. The reader is referred to \cite{pater2000} for more details.
\begin{example}
Consider \textnormal{RM}$_2(2,3)$, generated by vectors corresponding to the monomials of degree at most $2$ in variables
$x_0$, $x_1$ and $x_2$. The generator matrix $\text{G}$ of \textnormal{RM}$_2(2,3)$ is given as follows.
\begin{equation}\nonumber
\begin{bmatrix}
 11111111\\
 01010101\\
 00110011\\
 00001111\\
 00010001\\
 00000101\\
 00000011
\end{bmatrix}
\quad
\begin{matrix}
 1\\
 x_0\\
 x_1\\
 x_2\\
 x_0x_1\\
 x_0x_2\\
 x_1x_2
\end{matrix}
\end{equation}
\end{example}
\subsection{Quadratic Forms and Graphs of GBFs}
In this subsection, we introduce some lemmas and notations which will be used for our proposed constructions in the next section.
\begin{definition}
 Let $f$ be a GBF of $m$ variables $x_0, x_1, \cdots, x_{m-1}$ over $\mathbb{Z}_q$. Consider a list of $k$ $(0\leq k<m)$
 indices $0\leq j_0<j_1<\cdots j_k<m$ and write $\textbf{x}=(x_{j_0},x_{j_1},\cdots, x_{j_{k-1}})$. Also, consider  $\textbf{c}=(c_0,c_1,\cdots,c_{k-1})$ which is a fixed
 binary sequence. Define $\psi(f\arrowvert_{\textbf{x}=\textbf{c}})$ as a complex-valued sequence with
 $\omega^{f(i_0,i_1,\cdots, i_{m-1})}$ as $i$th component if $i_{j_\alpha}=c_{\alpha}$ for each
 $0\leq \alpha <k$ and equal to zero otherwise, where $\omega$ is a (complex-valued) $q$th root of unity. For $k=0$, $\psi(f\arrowvert_{\textbf{x}=\textbf{c}})$ reduces to
  the sequence $\psi(f)$ which has been defined in Subsection II-B.
\end{definition}
Let $Q$ be the quadratic form of $f$. Then, the GBF $f$ can be expressed as \cite{arthina}
\begin{equation}\label{Bool}
f=Q+\displaystyle \sum_{i=0}^{m-1}g_ix_i+g',
\end{equation}
where $g',g_i\in \mathbb{Z}_q$ are arbitrary.

 For a quadratic GBF $f$, let $G(f)$ denote the graph of $f$ which is obtained by joining the vertices $x_i$ and $x_j$ by an edge if there is
 a term $q_{i,j}x_ix_j$ $(0\leq i<j\leq m-1)$ in the GBF $f$ with $q_{i,j}\neq 0$ $(q_{i,j}\in \mathbb{Z}_q)$. Consider the function $f\arrowvert_{x_j=c}$, obtained
by substituting $x_j=c$ in $f$. It follows that the graph of $f\arrowvert_{x_j=c}$ is equal to the graph
obtained by deleting vertex $j$ from $G(f)$. Similarly the graph of $f\arrowvert_{\textbf{x}=\textbf{c}}$ is obtained by
deleting vertices $x_{j_0},x_{j_1},\cdots,x_{j_{k-1}}$ from $G(f)$. The final graph is independent of the choice of $\textbf{c}$. That is, for any
$\textbf{c}$, the quadratic part of the function $f\arrowvert_{\textbf{x}=\textbf{c}}$ is completely described by
the graph which is obtained from $G(f)$ by deleting vertices $x_{j_0},x_{j_1},\cdots,x_{j_{k-1}}$. Note that the quadratic forms in the
functions $f$ and $\tilde{f}$  are the same and therefore, they have the same associated graph.
 \begin{example}
  Let $f$ be a GBF of $3$ variables over $\mathbb{Z}_2$, as follows
  \begin{equation}\nonumber
 f(x_0, x_1, x_2)  =x_0x_2+x_2x_1+x_1+x_2.
  \end{equation}
  Let $\textbf{x}=(x_0,x_2)$ and $\textbf{c}=(0,1)$. Then, the complex-valued vector corresponding to $f\arrowvert_{x=c}$ can be written as follows.
\begin{equation}\nonumber
   \psi(f\arrowvert_{\textbf{x}=\textbf{c}})=(0,0,0,0,-,0,-,0).
  \end{equation}
 \end{example}
\subsection{Truncated Restricted Vectors}
Let $\psi(f)=(F_0,F_1,\cdots,F_{L-1})$ be a complex-valued vector of length $L$ and $\psi(f\arrowvert_{\textbf{x}=\textbf{c}})$ be a restriction of it.
Also, let  $i_{\textbf{c}}$ and $\bar{i_{\textbf{c}}}$ be the first and last nonzero entries in the restricted vector
$\psi(f\arrowvert_{\textbf{x}=\textbf{c}})$, we have
\begin{equation}\nonumber
 \psi(f\arrowvert_{\textbf{x}=\textbf{c}})=(0,\cdots,0,F_{i_{\textbf{c}}},F_{i_{\textbf{c}}+1},\cdots,F_{\bar{i_\textbf{c}}-1},F_{\bar{i_\textbf{c}}},0,\cdots,0),
\end{equation}
where the entries $F_i$ for $i_{\textbf{c}}<i<\bar{i_{\textbf{c}}}$, may not necessarily be nonzero. Then, the truncated vector is obtained by truncating
the leading and trailing zeros of the restricted vector which is denoted as follows. \\
\begin{equation}
 [\psi(f\arrowvert_{\textbf{x}=\textbf{c}})]=(F_{i_{\textbf{c}}},F_{i_{\textbf{c}}+1},\cdots,F_{\bar{i_\textbf{c}}-1},F_{\bar{i_\textbf{c}}}).
\end{equation}
\begin{lemma}\label{lemma1}\cite{pater2000}
Let $f,g$ be GBFs of $m$ variables.
Consider $0\leq j_0<j_1<\cdots<j_{k-1}<m$, which is a list of $k$ indices and $\textbf{c}=(c_0c_1\cdots c_{k-1})$
and $\textbf{d}=(d_0 d_1\cdots d_{k-1})$ are two binary vectors. Write $\textbf{x}=(x_{j_0}x_{j_1}\cdots x_{j_{k-1}})$
and consider $0\leq i_0<i_1<\cdots< i_{l-1}<m$, which is a set of indices which has no intersection with  $\{j_0,j_1,\cdots,j_{k-1}\}$.
Let $\textbf{y}=(x_{i_0}x_{i_1}\cdots x_{i_{l-1}})$, then
\begin{equation}
\begin{split}
C&\left (\psi(f\arrowvert_{\textbf{x}=\textbf{c}}),\psi(g\arrowvert_{\textbf{x}=\textbf{d}})\right )(\tau)\\&=\displaystyle\sum_{\textbf{c}_1,\textbf{c}_2}
C\left(\psi(f\arrowvert_{\textbf{xy}=\textbf{cc}_1}),\psi(g\arrowvert_{\textbf{xy}=\textbf{dc}_2})\right)(\tau).
\end{split}
\end{equation}
\end{lemma}
\begin{lemma}(\cite[Lemma. 1.20]{stinch})\label{lemma2}
Let $f$, $g$ be GBFs of $m$ variables and
$\psi(f\arrowvert_{\textbf{x}=\textbf{c}_1})$, $\psi(g\arrowvert_{\textbf{x}=\textbf{c}_2})$ be the corresponding vectors restricting  variables $\textbf{x}$ to $\textbf{c}_1$ and $\textbf{c}_2$ for $f$ and $g$, respectively.
Consider $i_{\textbf{c}_j}$ which is the index of the first nonzero entry in the vector $\psi(. \arrowvert_{\textbf{x}={\textbf{c}_j}})$, $j=1$ or $2$,
and $n_x$ the length of nonzero pattern. Then, the cross-correlation of the restricted vectors is given by a shifted cross-correlation of
the truncated vectors as follows.\\
$C\left(\psi(f\arrowvert_{\textbf{x}=\textbf{c}_1}),\psi(g\arrowvert_{\textbf{x}=\textbf{c}_2})\right)(\tau)$
 \begin{equation}\label{stinchcombelemma1}
=
\begin{cases}
C\left([\psi(f\arrowvert_{\textbf{x}=\textbf{c}_1})],[\psi(g\arrowvert_{\textbf{x}=\textbf{c}_2})]\right)(\tau-(i_{\textbf{c}_1}-i_{\textbf{c}_2})), & \\
~~~\text{if}~(i_{\textbf{c}_1}-i_{\textbf{c}_2})\!-\!(n_{\textbf{x}}-1)\leq \tau\leq  (i_{\textbf{c}_1}-i_{\textbf{c}_2})\!+\!(n_{\textbf{x}}-1); & \\
0,~\textnormal{otherwise}.&
\end{cases}
 \end{equation}
In particular, when $f=g$ and $\textbf{c}_1=\textbf{c}_2=\textbf{c}$,
\begin{equation}\nonumber
\begin{split}
 A&\left(\psi(f\arrowvert_{\textbf{x}=\textbf{c}})\right)(\tau)\\&=
\begin{cases}
A([\psi(f\arrowvert_{\textbf{x}=\textbf{c}})])(\tau), & -(n_{\textbf{x}}-1)\leq \tau\leq  (n_{\textbf{x}}-1), \\
0, & \textnormal{otherwise}.
\end{cases}
\end{split}
 \end{equation}
 \end{lemma}
 \textit{Lemma} \ref{lemma2}  will be used in the proof of \textit{Lemma} \ref{lemma6}.
\begin{example}
Let $f$ and $g$ be GBFs of $4$ variables over $\mathbb{Z}_2$, as follows.
\begin{equation}\nonumber
\begin{split}
 f(x_0,x_1,x_2,x_3)&=x_0x_1+x_2x_3+x_0,\\
 g(x_0,x_1,x_2,x_3)&=x_0x_2+x_1x_3.
 \end{split}
\end{equation}
Let $\textbf{x}=x_0x_2$, $\textbf{c}_1=(0,1)$ and $\textbf{c}_2=(1,0)$. Then, the restricted vectors of $f$ and $g$ at $\textbf{x}=\textbf{c}_1$ and
$\textbf{x}=\textbf{c}_2$ are
\begin{equation}\nonumber
\begin{split}
 \psi(f\arrowvert_{\textbf{x}=\textbf{c}_1})&=(0, 0, 0, 0, +, 0, +, 0, 0, 0, 0, 0, -, 0, -, 0),\\
 \psi(g\arrowvert_{\textbf{x}=\textbf{c}_2})&=(0, +, 0, +, 0, 0, 0, 0, 0, +, 0, -, 0, 0, 0, 0).
 \end{split}
\end{equation}
The length of the nonzero pattern $n_{\textbf{x}}=\bar{i_{\textbf{c}_j}}-i_{\textbf{c}_j}+1$, where $i_{\textbf{c}_j}$ and $\bar{i_{\textbf{c}_j}}$  are the indices of the first and last
nonzero entries in the vector $\psi(f\arrowvert_{\textbf{x}=\textbf{c}_j})$ $(j=1,2)$. Therefore $i_{\textbf{c}_1}=4$, $i_{\textbf{c}_2}=1$ and
$n_{\textbf{x}}=11$. Now, we have
\begin{equation}
\begin{split}
 \left (C\right.&\left.(\psi(f\arrowvert_{\textbf{x}=\textbf{c}_1}),\psi(g\arrowvert_{\textbf{x}=\textbf{c}_2}))(\tau) \right)_{\tau=-15}^{15}\\&=(0^8,
 -1,0^3,1,0^3,2,0,2,0^5,-1,0,-2,0,-1,0^2 ),
\end{split}
\end{equation}
where $0^m$ denotes $m$ consecutive zeros.

Therefore
\begin{equation}\label{stinchcombelemma2}
\begin{split}
 C&\left(\psi(f\arrowvert_{\textbf{x}=\textbf{c}_1}),\psi(g\arrowvert_{\textbf{x}=\textbf{c}_2})\right)(\tau)\\&=
\begin{cases}
C([\psi(f\arrowvert_{\textbf{x}=\textbf{c}_1})],[\psi(g\arrowvert_{\textbf{x}=\textbf{c}_2})])(\tau+3), & -7\leq \tau\leq  13, \\
0, & \textnormal{otherwise}.
\end{cases}
\end{split}
 \end{equation}
\end{example}

\begin{lemma}{(Construction of CCC \cite{arthina})}\label{lemma3}\\
Let $f$ be a GBF of $m$ variables and $\tilde{f}$ be its reversal. Suppose
$G(f)$ contains a set of $k$ distinct vertices labeled $j_0,j_1,\cdots,j_{k-1}$ with the property
that deleting those $k$  vertices and all their edges results in a path with $q/2$ being the weight of every edge of the path.
Let $(t_0,t_1,\cdots, t_{k-1})$ be the binary representation of the integer $t$. Define the complementary code $C_t$ to be
\begin{equation}
 \displaystyle \left\{ f\!+\!
 \frac{q}{2}\left(\displaystyle{\sum_{\alpha=0}^{k-1}} d_{\alpha}x_{j_{\alpha}}\!+\!\displaystyle{\sum_{\alpha=0}^{k-1}} t_{\alpha}x_{j_{\alpha}}
 \!+\!dx_{\gamma}\right): d,d_\alpha\in\{0,1\}     \right\},
\end{equation}
and the counterpart CC $\bar{C}_{2^k+t}$ to be
\begin{equation}
\displaystyle \left\{ \tilde{f}\!+\!
 \frac{q}{2}\left(\displaystyle{\sum_{\alpha=0}^{k-1}} d_{\alpha}\bar{x}_{j_{\alpha}}\!+\!\displaystyle{\sum_{\alpha=0}^{k-1}} t_{\alpha}\bar{x}_{j_{\alpha}}
 \!+\!\bar{d}x_{\gamma}\right): d,d_\alpha\in\{0,1\}     \right\},
 \end{equation}
 where $\gamma$ be the label of either end vertex in the path.
 Then
 \begin{equation}
  \{\psi(C_t):0\leq t<2^k\}\cup \{\psi^*(\bar{C}_{2^k+t}):0\leq t<2^k\}
 \end{equation}
 generate a set of CCC, where $\psi^*(\cdot)$ denotes the complex conjugate of $\psi(\cdot)$.
\end{lemma}
\textit{Lemma} \ref{lemma3} will be used in \textit{Theorem} \ref{theorem2} to show that the construction of CCC in \cite{arthina} is a special case
of our construction.
\begin{lemma} (\cite{liu2011})\label{lemma5}
 For any ZCCS with the parameters $K$, $M$, $L$ and $Z$, the theoretical bound is given by
\begin{equation}\label{op}
  K\leq M\lfloor L/Z\rfloor,
  \end{equation}
  where $Z$ is the ZCZ width, $K$ is the number of $Z$-complementary codes, $M$ is the number of constituent sequences in a $Z$-complementary code and
  $L$ is the length of each constituent sequence.
We call a ZCCS optimal if the equality in (\ref{op}) is achieved.
\end{lemma}
\section{Proposed Construction of Z-complementary code Set}
\label{sec:ZCS}
In this section, we present a direct construction of ZCCS over $\mathbb{Z}_q$ using a generic graph as shown in Fig. 1.
\begin{figure}[!t]
\centering
\includegraphics[height=9cm]{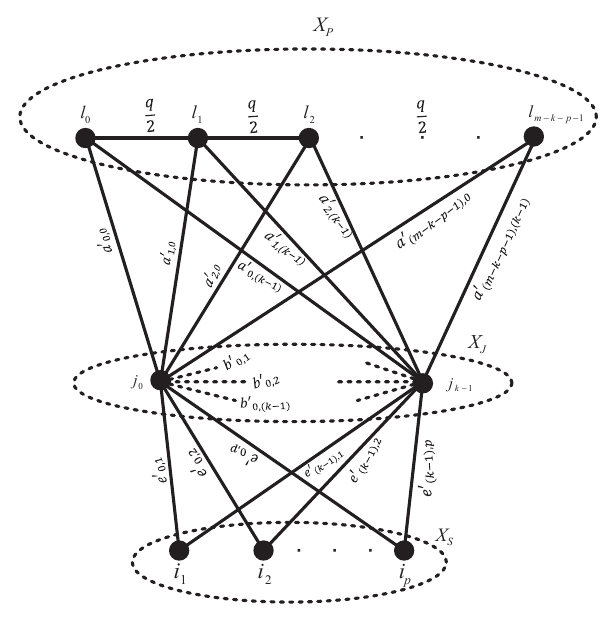}
\caption{The graph of the quadratic form $Q$.}
\end{figure}
Specifically, Fig. 1 contains $m$ vertices denoted by set $X_I=\{x_0,x_1,\cdots,x_{m-1}\}$. These $m$ vertices are divided  into
three disjoint sets: $X_{P}=\{x_{l_0},x_{l_1},\cdots,x_{l_{m-k-p-1}}\}$ represents
the vertices of a path whose edges have identical weight of $q/2$, $X_J=\{x_{j_0},x_{j_1},\cdots,x_{j_{k-1}}\}$, and
$X_S=\{x_{i_1},x_{i_2},$ $\cdots,x_{i_p}\}$. $a'_{i,\alpha}$'s denote the weights of the edges between
vertices from $X_P$ and $X_J$, $e'_{\alpha,\beta}$'s denote the weights of the edges between
vertices from $X_J$ and $X_S$, and $b'_{\alpha_1,\alpha_2}$'s denote the weights of edges between any two vertices from $X_{J}$.
From the graph, it is clear that after deleting all the vertices from the set $X_J$, the resulting graph
contains a path of vertices in the set $X_P$ and $p$ isolated vertices of the set $X_S$. The quadratic part of the GBF corresponding to the above
graph can be expressed as follows.
\begin{equation}
\begin{split}
Q=&\frac{q}{2}\sum_{i=0}^{m-k-p-2}x_{l_i}x_{l_{i+1}}+\sum_{i=0}^{m-k-p-1}\sum_{\alpha=0}^{k-1}a'_{i,\alpha}x_{l_i}x_{j_\alpha}\\&+\sum_{\alpha=0}^{k-1}\sum_{\beta=1}^{p}e'_{\alpha,\beta}x_{j_\alpha}x_{i_\beta}
 +\sum_{0\leq \alpha_1<\alpha_2<k}b'_{\alpha_1,\alpha_2}x_{j_{\alpha_1}}x_{j_{\alpha_2}},
 \end{split}
\end{equation}
where
$a'_{i,\alpha}$, $b'_{\alpha_1,\alpha_2}$ and $e'_{\alpha,\beta}\in \mathbb{Z}_q$.
We also need to define the following vectors which will be used throughout in our construction:
\begin{itemize}
 \item $\textbf{x}=(x_{j_0},x_{j_{1}},\cdots, x_{j_{k-1}})\in$ $\mathbb{Z}_2^k$,\\ $\textbf{x}'=(x_{i_1},x_{i_1},\cdots,x_{i_{p}})\in$ $\mathbb{Z}_2^p$.
 \item $\textbf{c}=(c_0,c_1,\cdots,c_{k-1})$, $\textbf{c}_i=(c_{i,0},c_{i,1},\cdots,c_{i,{k-1}})\in$ $\mathbb{Z}_2^k$.
 \item $\textbf{c}'=(c_1',c_2',\cdots,c_{p}')$, $\textbf{c}''=(c_1'',c_1'',\cdots,c_{p}'')$ \\and $\textbf{c}_j'=(c'_{j,1},c'_{j,2},\cdots,c'_{j,{p}})\in$ $\mathbb{Z}_2^p$.
\item $\textbf{d}'=(d'_1,d'_2,\cdots,d'_p)$, $\textbf{d}''=(d_1'',d_2'',\cdots,d_p'')\in$ $\mathbb{Z}_2^p$.
\item $\varGamma=(g_{i_1},g_{i_2},\cdots,g_{i_p})\in \mathbb{Z}_q^p$.
\end{itemize}
For ease of presentation, whenever the context is clear, we sometimes use $C(f,g)(\tau)$ to denote  $C(\psi(f),\psi(g))(\tau)$  for any two GBFs $f$ and $g$. Similar changes will be applied to
restricted Boolean functions also.

Furthermore, we need to define the following sets before presenting \textit{Theorem} \ref{theorem1}. Let
\begin{equation}\label{tau1}
T_{\textbf{c}_1'-\textbf{c}_2'}=\textbf{(c}_1'-\textbf{c}_2')\cdot (2^{i_1},2^{i_2},\cdots, 2^{i_{p}}),
\end{equation}
and
\begin{equation}\label{tau2}
R_{\tau_{i}}=\{(\textbf{c}_1',\textbf{c}_2'):T_{\textbf{c}_1'-\textbf{c}_2'}=\tau_i, \textbf{c}_1'\neq\textbf{c}_2',\textbf{c}_1',\textbf{c}_2'
\in \mathbb{Z}^p_2\}.
\end{equation}
Here $T_{\textbf{c}_1'- \textbf{c}_2'}$ represents nonzero time-shift for a binary pair of vectors $(\textbf{c}_1',\textbf{c}_2')$.
$R_{\tau_{i}} $ is a set which contains all pairs $(\textbf{c}_1',\textbf{c}_2')$ such that
$T_{\textbf{c}_1'-\textbf{c}_2'}=\tau_i$. From (\ref{tau1}) and (\ref{tau2}), it is observed that the number of such
distinct $\tau_i$ is at most $3^p-1$.
\begin{example}\label{exafine}
Let $p=2$, $i_1=2$ and $i_2=3$. Then, we have $T_{\textbf{c}_1'-\textbf{c}_2'}=\textbf{(c}_1'-\textbf{c}_2')\cdot (2^{2},2^{3})$
whose values are taken from the set $\{\pm4,\pm8,\pm12\}$ depending on $\textbf{c}'_1,\textbf{c}'_2$. $R_{\tau_{i}}=\{(\textbf{c}_1',\textbf{c}_2'):T_{\textbf{c}_1'-\textbf{c}_2'}=\tau_i, \textbf{c}_1'\neq\textbf{c}_2',\textbf{c}_1',\textbf{c}_2'
\in \mathbb{Z}^2_2\}$ for $\tau_i=4,-4,8,-8,12,-12$ are given below.
\begin{equation}\nonumber
\begin{split}
R_{4}&=\{((0,1),(1,0)),((1,1),(0,1)),((1,0),(0,0))\},\\
R_{-4}&=\{((0,0),(1,0)),((0,1),(1,1)),((1,0),(0,1))\},\\
R_{8}&=\{((0,1),(0,0)),((1,1),(1,0))\},\\
R_{-8}&=\{((0,0),(0,1)),((1,0),(1,1))\},\\
R_{12}&=\{((1,1),(0,0))\},\\
R_{-12}&=\{((0,0),(1,1))\}.
\end{split}
\end{equation}
\end{example}
\subsection{Construction of Z-complementary Code Set}
In the above context, we are ready to present a construction of ZCCS over $\mathbb{Z}_q$.
Let $f$ be a GBF with $m$ variables
and $Q$ be the quadratic part of $f$. For $0\leq t \leq 2^{k+p}-1$, define the order set $S_t$, as follows.
\begin{equation}\label{ncs}
\begin{split}
 \displaystyle\left \{Q+
 \displaystyle{\sum_{i=0}^{m-1}}g_ix_i+g'+
 \frac{q}{2}\left(\displaystyle{\sum_{\alpha=0}^{k-1}} d_{\alpha}x_{j_{\alpha}}+\displaystyle{\sum_{\alpha=0}^{k-1}} b_{\alpha}x_{j_{\alpha}}\right.\right.\\ \left.\left.+
 \displaystyle{\sum_{\alpha=1}^{p}} d'_{\alpha}x_{i_{\alpha}}+dx_{\gamma}\right): d,d_\alpha\in \mathbb{Z}_2     \right\},
\end{split}
\end{equation}
where $t=\displaystyle\sum_{\alpha =0}^{k-1}b_{\alpha} 2^{\alpha}+\sum_{\alpha =k}^{k+p-1}d'_{\alpha-k+1} 2^{\alpha}$.

Let $\textbf{d}=(d_0,d_1,\cdots, d_{k-1})$, $\textbf{b}=(b_0,b_1,\cdots b_{k-1})$, $\textbf{b}'=(b'_0,b'_1,\cdots, b'_{k-1})$ ($b_i,b_i',d_i\in\{0,1\}$, $i=0,1,\cdots,k-1$) be binary vectors and
$\textbf{bd}'= (b_0,b_1,\cdots, b_{k-1},d'_1,d'_2,\cdots, d'_{p})$, $\textbf{b}'\textbf{d}''=(b'_0,b'_1,\cdots, b'_{k-1},d''_1,d''_2,\cdots, d''_{p})$ are binary
representations of $t$, $t'$ $(0\leq t,t'\leq 2^{k+p}-1)$ respectively, where $t'=\displaystyle\sum_{\alpha =0}^{k-1}b'_{\alpha} 2^{\alpha}+\sum_{\alpha =k}^{k+p-1}d''_{\alpha-k+1} 2^{\alpha}$. In (\ref{ncs}), $g_0,g_1,\cdots,g_{m-1}$ are the coefficients of $x_0,x_1,\cdots,x_{m-1}$.
In the beginning of this Section, we have defined $\varGamma=(g_{i_1},g_{i_2},\cdots,g_{i_p})$ ($p<m$), where $i_1,i_2,\cdots,i_p$ all are distinct and belong to the set $\{0,1,\cdots,m-1\}$. In another words, we can say that $g_{i_1},g_{i_2},\cdots,g_{i_p}$ are the coefficients of $x_{i_1},x_{i_2},\cdots,x_{i_p}$. Therefore,
the term $\displaystyle\sum_{i=0}^{m-1}g_ix_i$ presented in (\ref{ncs}), can be expressed as
\begin{equation}
 \begin{split}
  \displaystyle\sum_{i=0}^{m-1}g_ix_i&=\sum_{i\in\{0,1,\!\cdots\!,m-1\}\!\setminus\!\{i_1,i_2,\!\cdots\!,i_p\}}g_ix_i+\left(g_{i_1}x_{i_1}+g_{i_2}x_{i_2}\right. \\ &\left.\qquad\qquad\qquad\qquad\qquad\qquad+\cdots+g_{i_p}x_{i_p}\right)\\
  &=\sum_{i\in\{0,1,\!\cdots\!,m-1\}\!\setminus\!\{i_1,i_2,\!\cdots\!,i_p\}}g_ix_i+\textbf{x}'\cdot\varGamma,
 \end{split}
\end{equation}
where $\textbf{x}'\cdot\varGamma=g_{i_1}x_{i_1}+g_{i_2}x_{i_2}+\cdots+g_{i_p}x_{i_p}$.
\begin{theorem}\label{theorem1}
Suppose $G(f)$ satisfies the property
that deleting $k$ vertices specified in $X_J$ and all their associated edges results in a path and $p$ isolates
vertices in $X_S$. Let $\gamma$ be the label of
either end vertex in the path. Then for any choice of $g',g_i \in \mathbb{Z}_q$, the auto-correlation function of the code $\psi(S_t)$
and the cross-correlation function between two codes $\psi(S_t)$ and $\psi(S_{t'})$  are as follows.
\end{theorem}
$1)$ For $ \textbf{b}'=\textbf{b},\textbf{d}'= \textbf{d}''$
\begin{equation}\label{auto1}
\begin{split}
A&(\psi(S_t))(\tau)\\&=
\begin{cases}
2^{m+k+1}, ~~~~~~~~~~~~~~~~~~~~~~~~\tau =0,\\
2^{m-p+1}\displaystyle\sum_{(\textbf{c}',\textbf{c}'')\in R_{\tau}}
(-1)^{\textbf{d}'\cdot(\textbf{c}'+\textbf{c}'')}\omega^{(\textbf{c}'-\textbf{c}'')\cdot \varGamma}\\~~~~~~~~~~~~\times\sum_{\textbf{c}}\omega^{g_{\textbf{c}\textbf{c}'}-g_{\textbf{c}\textbf{c}''}},~~\tau=\tau_i, i=1,2,\cdots,r,\\
0, ~~~~~~~~~~~~~~~~~~~~~~~~~~~~~~~~~~\text{otherwise}.
\end{cases}
\end{split}
\end{equation}
$2)$
\begin{equation}\label{cross1}
\begin{split}
 C&(\psi(S_t),\psi(S_{t'}))(\tau)\\&=
\begin{cases}
2^{m-p+1}\displaystyle\sum_{(\textbf{c}',\textbf{c}'')\in R_{\tau}}(-1)^{\textbf{d}'\cdot\textbf{c}'+\textbf{d}''\cdot\textbf{c}''}
\omega^{(\textbf{c}'-\textbf{c}'')\cdot \varGamma}\\\!\!\times\!\!\displaystyle\left (\sum_{\textbf{c}}\omega^{g_{\textbf{c}\textbf{c}'}-g_{\textbf{c}\textbf{c}''}}(-1)^{(\textbf{b}+\textbf{b}')\cdot\textbf{c}}\right),
\tau\!=\!\tau_i, i\!=\!1,2,\cdots,r,\\
0, ~~~~~~~~~~~~~~~~~~~~~~~~~~~~~~~~~~~~~\text{otherwise},
\end{cases}\\
\end{split}
\end{equation}
where $0\leq r \leq 3^p-1$, $g_{\textbf{c}\textbf{c}'}=\sum_{\alpha=0}^{k-1}\sum_{\beta=1}^{p}e'_{\alpha,\beta}c_{\alpha}c_{\beta}'$ and $ g_{\textbf{c}\textbf{c}''}=\sum_{\alpha=0}^{k-1}\sum_{\beta=1}^{p}e'_{\alpha,\beta}c_{\alpha}c_{\beta}''$.
\begin{IEEEproof}
See Appendix A.
\end{IEEEproof}
\begin{corollary}\label{coro1}
In the context of \textit{Theorem} \ref{theorem1}, consider $m-p, m-p+1,\cdots, m-1$, as the labels of $p$ isolated vertices. Then
 $\{\psi({S}_t): 0\leq t\leq 2^{k+p}-1\}$ is a $(2^{k+p},2^{m-p})$-$\text{ZCCS}_{2^{k+1}}^{2^m}$.
\end{corollary}
\begin{IEEEproof}
Let $s=\min\{\left |\tau_i \right |: i=1,2,\cdots,r  \}$, where \\$\tau_i=\textbf{(c}_1'-\textbf{c}_2')\cdot (2^{m-p},2^{m-p+1},\cdots, 2^{m-1})$.\\
To find $s$, we start with
\begin{equation}\label{pur}
\begin{split}
\left |\tau_i\right |&=\left |\textbf{(c}_1'-\textbf{c}_2')\cdot (2^{m-p},2^{m-p+1},\cdots, 2^{m-1})\right |\qquad\qquad\qquad\qquad\qquad\qquad\qquad\qquad\\
&=\left |\displaystyle\sum_{j=1}^{p}(c'_{1,j}-c'_{2,j})2^{m-p+j-1}\right |\qquad\qquad\qquad\qquad\qquad\qquad\qquad\qquad\qquad\qquad\quad\\
&=2^{m-p}\Bigl |\{(c'_{1,1}-c'_{2,1})+(c'_{1,2}-c'_{2,2})2+\cdots \\&~~~~~~~~~~~~+(c'_{1,p}-c'_{2,p})2^{p-1}\}\Bigl |\\
&\geq 2^{m-p},\quad\textnormal{for}\quad \textbf{c}_1\neq \textbf{c}_2.\qquad\qquad\qquad\qquad\qquad\qquad\qquad\qquad\qquad\qquad\\
\end{split}
\end{equation}
Therefore, $\left |\tau_i\right |\geq 2^{m-p} $ $\forall i=1,2,\cdots,r$, where
the equality is met if $c'_{1,j}=c'_{2,j} $ for all $j$, except for $j=1$.
Hence,
\begin{equation}\label{min_isol}
s=\min\{\left |\tau_i\right |: i=1,2,\cdots,r  \}=2^{m-p}.
\end{equation}
From \textit{Theorem} \ref{theorem1} and (\ref{min_isol}), it is asserted that for any $t,t'$ $(0\leq t,t'\leq 2^{k+p}-1)$
\begin{equation}
C(\psi(S_t),\psi(S_{t'}))(\tau) =
\begin{cases}
0, & 0<\left |\tau\right |<2^{m-p},  t=t',\\
0, & \left |\tau\right |\leq 2^{m-p},  t\neq t'.
\end{cases}
\end{equation}
Therefore the set $\{\psi(S_t): 0\leq t\leq 2^{k+p}-1\}$ is a $(2^{k+p},2^{m-p})$-$\text{ZCCS}_{2^{k+1}}^{2^m}$.
\end{IEEEproof}
For each $0\leq t \leq 2^{k+p}-1$, define the order set $\bar{S}_t$ as follows.
\begin{equation}
\begin{split}
 \displaystyle\left \{\tilde{f}+
 \frac{q}{2}\left(\displaystyle{\sum_{\alpha=0}^{k-1}} d_{\alpha}\bar{x}_{j_{\alpha}}+\displaystyle{\sum_{\alpha=0}^{k-1}}
 b_{\alpha}\bar{x}_{j_{\alpha}}+\right.\right.\qquad\qquad\qquad\qquad \\ \qquad\qquad\qquad\qquad\left. \left. \displaystyle{\sum_{\alpha=0}^{p-1}} d'_{\alpha}\bar{x}_{i_{\alpha}}+\bar{d}x_{\gamma}\right): d,d_\alpha\in\{0,1\}\right\}.
\end{split}
\end{equation}
\begin{corollary}\label{coro2}
In the context of Theorem $1$, consider $m-p, m-p+1,\cdots, m-1$, as the labels of $p$ isolated vertices. Then
$\{\psi(\bar{S}_t): 0\leq t\leq 2^{k+p}-1\}$ is a $(2^{k+p},2^{m-p})$-$\text{ZCCS}_{2^{k+1}}^{2^m}$.
\end{corollary}
The proof of \textit{Corollary} \ref{coro2} follows directly from the proofs of \textit{Theorem} \ref{theorem1} and \textit{Corollary} \ref{coro1}.
\begin{theorem}\label{theorem2}
 Consider $\{\psi(S_t)\}$ and  $\{\psi(\bar{S}_t\})$ in \textit{Corollary} \ref{coro1} and \textit{Corollary} \ref{coro2}, respectively. Then,
 \begin{equation}\nonumber
 \{\psi({S}_t): 0\leq t\leq 2^{k+p}-1\}\cup \{\psi^*(\bar{S}_t): 0\leq t\leq 2^{k+p}-1\},
 \end{equation}
 form $(2^{k+p+1},2^{m-p})$-$\text{ZCCS}_{2^{k+1}}^{2^m}$.
\end{theorem}
\begin{IEEEproof}
See Appendix B.
\end{IEEEproof}
It is noted that our proposed ZCCS is optimal with respect to the theoretical bound in \textit{Lemma} \ref{lemma5}. Also, when $p=0$, our proposed
construction in \textit{Theorem} \ref{theorem2}  reduces to that in \cite[Th.~3.6]{arthina}.
\begin{remark}
 From \textit{Theorem} \ref{theorem2} and Fig. 1, it is observed that at least
 \begin{displaymath}
 \frac{(m-p)!}{2(k!)}(q-1)^{k(m-k-p)}q^{kp+\frac{k(k-1)}{2}+m+1}
 \end{displaymath}
 distinct optimal ZCCSs can be constructed from our proposed construction.
\end{remark}
\begin{IEEEproof}
See Appendix C.
\end{IEEEproof}
\subsection{EXAMPLE}
\begin{figure}[!t]
\centering
\includegraphics[height=4cm]{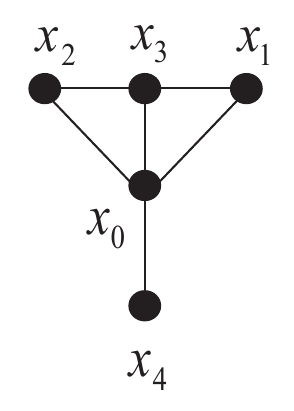}
\caption{The graph of the quadratic Boolean function  $x_2x_3+x_3x_1+x_0x_2+x_0x_3+x_0x_1+x_0x_4$.}
\end{figure}
In this subsection, we provide an example to illustrate our proposed ZCCS construction.
\begin{example}
Let $f$ be a GBF of $5$ variables over $\mathbb{Z}_4$, where the associated graph $G(f)$ is given in Fig. 2. Note that $G(f)$ is a
graph of five vertices satisfying the property that deleting $x_0$, the entire graph reduces to a path consisting of vertices labeled $2,3,1$
and an isolated vertex labeled $4$. This graph leads to an optimal ZCCS as follows. Let
\begin{equation}\nonumber
 Q=2(x_2x_3+x_3x_1+x_0x_2+x_0x_3+x_0x_1+x_0x_4),
\end{equation}
and
\begin{equation}\nonumber
 f(x_0,x_1,x_2,x_3,x_4)=Q+x_0+3x_1.
\end{equation}
Also, let
\begin{equation}\label{ab3}
S_t\!=\!\left\{f\!\!+\!\!2(d_0x_0\!+\!b_0x_0\!+\!d_0'x_4\!+\!dx_1)\!:\! d,d_0\!\in\! \mathbb{Z}_2\right\}, 0\leq \!t\! \leq 3,
\end{equation}
and
\begin{equation}\label{ab4}
\bar{S}_t\!=\!\left\{\tilde{f}\!\!+\!\!2(d_0\bar{x}_0\!+\!b_0\bar{x}_0\!+\!d_0'\bar{x}_4\!+\!\bar{d}x_1)\!:\! d,d_0\!\in\! \mathbb{Z}_2\right\}, 0\leq\! t\! \leq 3,
\end{equation}
where $t=b_02^0+d_0'2^1$ $(b_0,d_0'\in \mathbb{Z}_2)$.
Consider $C_t=\psi(S_t)$ and $C_{2^2+t}=\psi^*(\bar{S}_t)$, given in Table I.
\begin{table}
\centering
\tiny
\begin{tabular}{ |c|c| }
 \hline
 \multicolumn{2}{|c|}{$(8,16)$-$\text{ZCCS}_{4}^{32}$} \\
 \hline
 $C_0$&$C_1$ \\
 \hline
 $0     1	3	2	0	3	3	0	0	3	1	2	2	3	3	2	0	3	3	0	0	1	3	2	0	1	1	0	2	1	3	0$&$0	3	3	0	0	1	3	2	0	1	1	0	2	1	3	0	0	1	3	2	0	3	3	0	0	3	1	2	2	3	3	2$\\
 $0	3	3	0	0	1	3	2	0	1	1	0	2	1	3	0	0	1	3	2	0	3	3	0	0	3	1	2	2	3	3	2$&$0	1	3	2	0	3	3	0	0	3	1	2	2	3	3	2	0	3	3	0	0	1	3	2	0	1	1	0	2	1	3	0$\\
 $0	1	1	0	0	3	1	2	0	3	3	0	2	3	1	0	0	3	1	2	0	1	1	0	0	1	3	2	2	1	1	2$&$0	3	1	2	0	1	1	0	0	1	3	2	2	1	1	2	0	1	1	0	0	3	1	2	0	3	3	0	2	3	1	0$\\
 $0	3	1	2	0	1	1	0	0	1	3	2	2	1	1	2	0	1	1	0	0	3	1	2	0	3	3	0	2	3	1	0$&$0	1	1	0	0	3	1	2	0	3	3	0	2	3	1	0	0	3	1	2	0	1	1	0	0	1	3	2	2	1	1	2$\\
 \hline
 $C_2$&$C_3$\\
 \hline
 $0	1	3	2	0	3	3	0	0	3	1	2	2	3	3	2	2	1	1	2	2	3	1	0	2	3	3	2	0	3	1	2$ & $0	3	3	0	0	1	3	2	0	1	1	0	2	1	3	0	2	3	1	0	2	1	1	2	2	1	3	0	0	1	1	0$ \\
 $0	3	3	0	0	1	3	2	0	1	1	0	2	1	3	0	2	3	1	0	2	1	1	2	2	1	3	0	0	1	1	0$ & $0	1	3	2	0	3	3	0	0	3	1	2	2	3	3	2	2	1	1	2	2	3	1	0	2	3	3	2	0	3	1	2$ \\
 $0	1	1	0	0	3	1	2	0	3	3	0	2	3	1	0	2	1	3	0	2	3	3	2	2	3	1	0	0	3	3	0$ & $0	3	1	2	0	1	1	0	0	1	3	2	2	1	1	2	2	3	3	2	2	1	3	0	2	1	1	2	0	1	3	2$  \\
 $0	3	1	2	0	1	1	0	0	1	3	2	2	1	1	2	2	3	3	2	2	1	3	0	2	1	1	2	0	1	3	2$ & $0	1	1	0	0	3	1	2	0	3	3	0	2	3	1	0	2	1	3	0	2	3	3	2	2	3	1	0	0	3	3	0$ \\
  \hline
  $C_4$&$C_5$\\
  \hline
  $0	1	1	0	0	3	1	2	2	1	1	2	0	1	3	2	2	1	3	0	2	3	3	2	0	1	3	2	2	1	1	2$ & $2	1	3	0	2	3	3	2	0	1	3	2	2	1	1	2	0	1	1	0	0	3	1	2	2	1	1	2	0	1	3	2$ \\
  $2	1	3	0	2	3	3	2	0	1	3	2	2	1	1	2	0	1	1	0	0	3	1	2	2	1	1	2	0	1	3	2$ & $0	1	1	0	0	3	1	2	2	1	1	2	0	1	3	2	2	1	3	0	2	3	3	2	0	1	3	2	2	1	1	2$\\
  $0	1	3	2	0	3	3	0	2	1	3	0	0	1	1	0	2	1	1	2	2	3	1	0	0	1	1	0	2	1	3	0$ & $2	1	1	2	2	3	1	0	0	1	1	0	2	1	3	0	0	1	3	2	0	3	3	0	2	1	3	0	0	1	1	0$\\
  $2	1	1	2	2	3	1	0	0	1	1	0	2	1	3	0	0	1	3	2	0	3	3	0	2	1	3	0	0	1	1	0$ & $0	1	3	2	0	3	3	0	2	1	3	0	0	1	1	0	2	1	1	2	2	3	1	0	0	1	1	0	2	1	3	0$ \\
  \hline
  $C_6$&$C_7$\\
  \hline
  $2	3	3	2	2	1	3	0	0	3	3	0	2	3	1	0	2	1	3	0	2	3	3	2	0	1	3	2	2	1	1	2$ & $0	3	1	2	0	1	1	0	2	3	1	0	0	3	3	0	0	1	1	0	0	3	1	2	2	1	1	2	0	1	3	2$ \\
  $0	3	1	2	0	1	1	0	2	3	1	0	0	3	3	0	0	1	1	0	0	3	1	2	2	1	1	2	0	1	3	2$ & $2	3	3	2	2	1	3	0	0	3	3	0	2	3	1	0	2	1	3	0	2	3	3	2	0	1	3	2	2	1	1	2$ \\
  $2	3	1	0	2	1	1	2	0	3	1	2	2	3	3	2	2	1	1	2	2	3	1	0	0	1	1	0	2	1	3	0$ & $0	3	3	0	0	1	3	2	2	3	3	2	0	3	1	2	0	1	3	2	0	3	3	0	2	1	3	0	0	1	1	0$ \\
  $0	3	3	0	0	1	3	2	2	3	3	2	0	3	1	2	0	1	3	2	0	3	3	0	2	1	3	0	0	1	1	0$ & $2	3	1	0	2	1	1	2	0	3	1	2	2	3	3	2	2	1	1	2	2	3	1	0	0	1	1	0	2	1	3	0$ \\
  \hline
\end{tabular}
\caption{Optimal ZCCS over the alphabet $\mathbb{Z}_4$.}
\end{table}
\begin{figure}
\centering
\includegraphics[height=10cm,width=9.5cm]{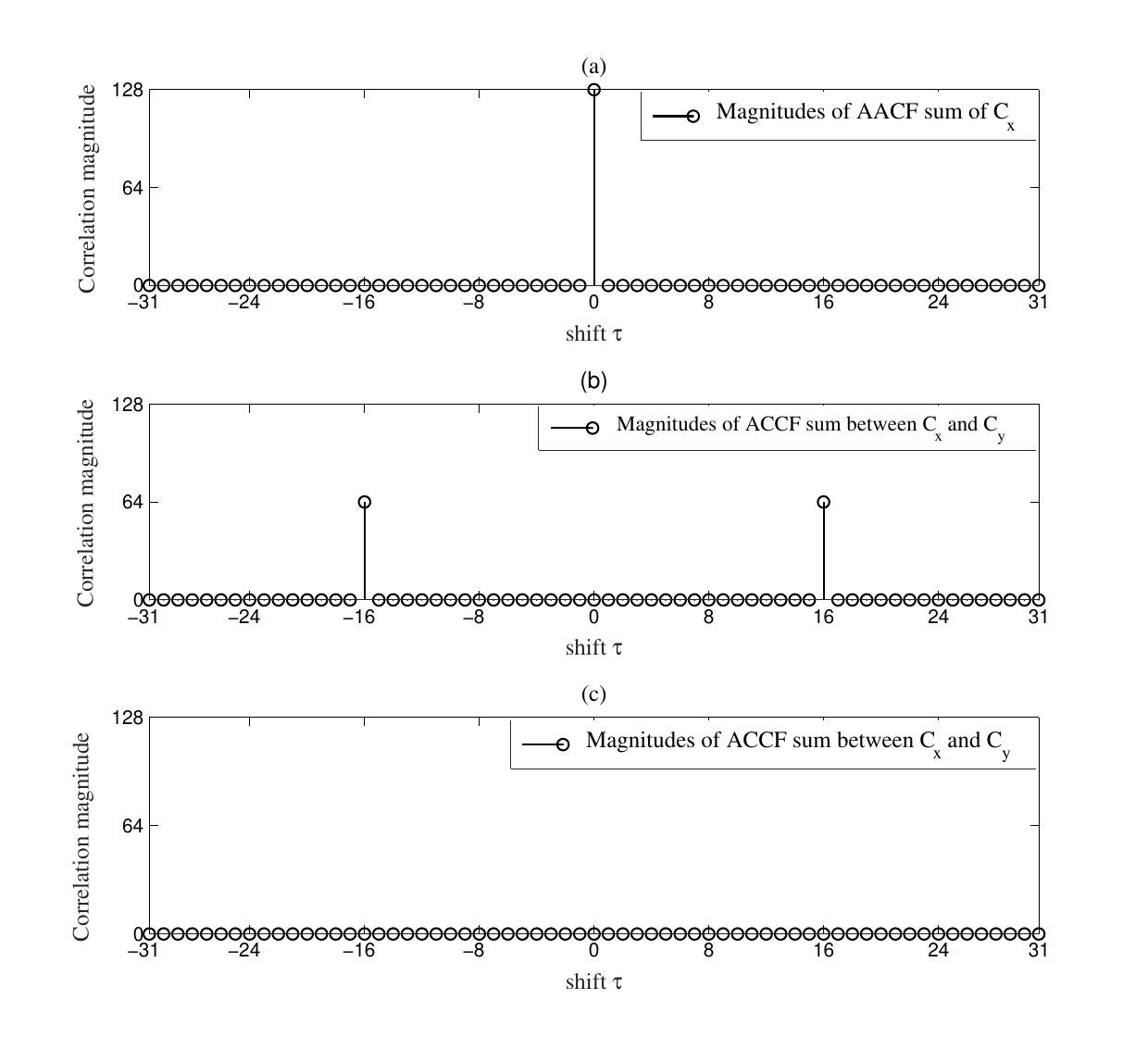}
\caption{Correlation plots of $(8,16)$-$\text{ZCCS}_{4}^{32}$ in Table I.}
\end{figure}
The correlation properties of the ZCCS in Table I are illustrated in Fig. 3. Specifically,
Fig. 3-a presents the absolute value of AACF sum of each code $C_x$ from $\{C_0,C_1,\cdots,C_7\}$, Fig. 3-b shows
absolute value of ACCF sum between any two distinct codes $C_x$ and $C_y$ ($x=b_02^0+d_0'2^1$, $y=b'_02^0+d''_02^1$ or, $x=2^2+b_02^0+d_0'2^1$, $y=2^2+b'_02^0+d''_02^1$) from
$\{C_0,C_1,C_2,C_3\}$ (or from $\{C_4,C_5,C_6,C_7\}$) with the condition $b_0\neq b_0'$. Fig. 3-c presents
the absolute value of ACCF sum between any two distinct codes $C_x$ and $C_y$ with the following senarios:
\begin{enumerate}
 \item The codes are drawn from $\{C_0,C_1,C_2,C_3\}$ (or from $\{C_4,C_5,C_6,C_7\}$) with the condition $b_0=b_0'$.
 \item One  code is drawn from $\{C_0,C_1,C_2,C_3\}$ and the other code from $\{C_4,C_5,C_6,C_7\}$.
\end{enumerate}
It is seen that the ZCZ width is $16$. Hence, the ZCCS satisfies the equality of (\ref{op}) as
$K=8$, $M=4$, $Z=16$ and $N=32$ and therefore, the ZCCS in Table I is optimal.
\end{example}
\section{CONCLUSION}
\label{sec:con}
In this paper, we have proposed
a direct construction of ZCCS using graphical representation of second-order RM codes. The proposed
construction valids for any number of isolated vertices present in the graph, is capable of generating
optimal ZCCS with respect to the set size upper bound in \textit{Lemma} \ref{lemma5}. It is noted that the construction of CCCs in \cite{arthina} is
a special case of our work work when the number of isolated vertices is set to zero. Flexible ZCZ
width and set size can be obtained by varying the number of isolated vertices.
\begin{appendices}
\section{Proof of \textnormal{\textit{Theorem} \ref{theorem1}}}
Before proving the \textit{Theorem} \ref{theorem1}, we present \textit{Lemma} \ref{lemma6} and \textit{Lemma} \ref{lemma4} where \textit{Lemma} \ref{lemma6} will be used in the proof of \textit{Theorem} \ref{theorem1} and
\textit{Lemma} \ref{lemma4} will be used in the proof of both \textit{Theorem} \ref{theorem1}
 and \textit{Theorem} \ref{theorem2}.
\begin{lemma}\label{lemma6}
Let $f$ and $f'$ be two GBFs of $m$ variables $x_0,x_1,\cdots,x_{m-1}$
$(m\geq 2)$, such that for some $k$ $(0\leq k\leq m-p-2$, $p\geq 0)$,  $f\arrowvert _{\textbf{x}=\textbf{c}}$
and $f'\arrowvert_{\textbf{x}=\textbf{c}}$ are given by
\begin{equation}\nonumber
\begin{split}
f\arrowvert _{\textbf{x}=\textbf{c}}&=P+L+g_{i_1}x_{i_1}+g_{i_2}x_{i_2}+\cdots+g_{i_p}x_{i_p}+g',\qquad\qquad \\
f'\arrowvert _{\textbf{x}=\textbf{c}}&=f\arrowvert _{\textbf{x}=\textbf{c}}+\frac{q}{2}x_{\gamma},
\end{split}
\end{equation}
where
\begin{equation}\nonumber
\begin{split}
P&=\frac{q}{2}\displaystyle\sum_{\alpha=0}^{m-k-p-2}x_{l_{\alpha}}x_{l_{\alpha+1}},\\
L&=
 \displaystyle\sum_{\alpha=0}^{m-k-p-1}g_{l_\alpha}x_{l_{\alpha}},
 \end{split}
\end{equation}$g_{l_\alpha},g'\in \mathbb{Z}_q$, $\alpha =0,1,\cdots,m-k-p-1$ and $\gamma$ is the label of  either
end vertex of the path $G(P)$.
Then for fixed $\textbf{c}$ and $\textbf{d}'\neq\textbf{d}''$, we have \\
 $C\left(f\arrowvert _{\textbf{x}\textbf{x}'=\textbf{c}\textbf{d}'},f\arrowvert _{\textbf{x}\textbf{x}'=\textbf{c}\textbf{d}''}\right)(\tau)+C\left(f'\arrowvert _{\textbf{x}\textbf{x}'=\textbf{c}\textbf{d}'},f'\arrowvert _{\textbf{x}\textbf{x}'=\textbf{c}\textbf{d}''}\right)(\tau)$
 \begin{equation}
 \begin{split}
  =\begin{cases}
\omega^{(d_1'-d_1'')g_{i_1}}+\cdots+(d_p'-d_p'')g_{i_p}2^{m-(k+p)+1},  \\~~~~~~~~~~~~~~\tau=(d_1'-d_1'')2^{i_1}+\cdots+(d_p'-d_p'')2^{i_p},\\
0, ~~~~~~~~~~~~\textnormal{otherwise}.
 \end{cases}
 \end{split}
 \end{equation}
\begin{IEEEproof}
Using \textit{Lemma} \ref{lemma2}, in terms of the truncated vectors, the sum of cross-correlations of the hypothesis becomes
\begin{equation}\nonumber
\begin{split}
C\left([f\arrowvert _{\textbf{x}\textbf{x}'=\textbf{c}\textbf{d}'}],
[f\arrowvert _{\textbf{x}\textbf{x}'=\textbf{c}\textbf{d}''}]\right)(\tau-(u_1-u_2))\qquad\qquad\qquad\\+C\left([f'\arrowvert _{\textbf{x}\textbf{x}'
=\textbf{c}\textbf{d}'}],[f'\arrowvert _{\textbf{x}\textbf{x}'=\textbf{c}\textbf{d}''}]\right)(\tau-(u_1-u_2)),\\
(u_1-u_2)-(n_x-1)\leq\tau\leq (u_1-u_2)+(n_x-1),
\end{split}
\end{equation}
where $u_1$ is the index of the first nonzero entry
in the  vector $\psi((.)\arrowvert_{\textbf{x}\textbf{x}'=\textbf{c}\textbf{d}'})$ and $u_2$ that in $\psi((.)\arrowvert_{\textbf{x}\textbf{x}'
=\textbf{c}\textbf{d}''})$.  For $\tau$ outside of this range,
each cross-correlation is zero by the \textit{Lemma} \ref{lemma2}, so the sum is zero too. For convenience write $\tau '=\tau-(u_1-u_2)$ and thus
we consider the sum as follows.
\begin{equation}\label{exp}
\begin{split}
 C\left([f\arrowvert _{\textbf{x}\textbf{x}'=\textbf{c}\textbf{d}'}],[f\arrowvert _{\textbf{x}\textbf{x}'=
 \textbf{c}\textbf{d}''}]\right)(\tau ')\qquad\qquad\qquad\qquad\qquad\qquad\qquad\\+C\left([f'\arrowvert _{\textbf{x}\textbf{x}'=\textbf{c}\textbf{d}'}],[f'
 \arrowvert _{\textbf{x}\textbf{x}'=\textbf{c}\textbf{d}''}]\right)(\tau '),-(n_x-1)\leq\!\!\tau'\!\!\leq (n_x-1).\quad
 \end{split}
 \end{equation}
Next we note that
\begin{equation}
f\arrowvert_{\textbf{x}\textbf{x}'=\textbf{c}\textbf{d}'}=f\arrowvert_{\textbf{x}\textbf{x}'=\textbf{c}\textbf{d}''}+
(d_1'-d_1'')g_{i_1}+\cdots+(d_p'-d_p'')g_{i_p},
\end{equation}
which means that the nonzero values in the vector $\psi(f\arrowvert_{\textbf{x}\textbf{x}'=\textbf{c}\textbf{d}'})$ are
$\omega^{(d_1'-d_1'')g_{i_1}+\cdots+(d_p'-d_p'')g_{i_p}}$ times those in the vector $\psi(f\arrowvert_{\textbf{x}\textbf{x}'
=\textbf{c}\textbf{d}''})$, only shifted relative to each other. For truncated vectors we have
\begin{equation}\label{sab1}
[\psi(f\arrowvert _{\textbf{x}\textbf{x}'=\textbf{c}\textbf{d}'})]
=\omega^{(d_1'-d_1'')g_{i_1}+\cdots+(d_p'-d_p'')g_{i_p}}[\psi(f\arrowvert _{\textbf{x}\textbf{x}'=\textbf{c}\textbf{d}''})].
\end{equation}
Substituting (\ref{sab1}) and its equivalent expression for $\psi(f'\arrowvert_{\textbf{xx}'=\textbf{cd}'})$ into (\ref{exp}), we have for all $\tau '$,
\begin{equation}\label{exp1}
\begin{split}
 &C([f\arrowvert _{\textbf{x}\textbf{x}'
 =\textbf{c}\textbf{d}'}],[f\arrowvert _{\textbf{x}\textbf{x}'=\textbf{c}\textbf{d}''}])(\tau ')\\&~~~~~~~~~~~~~~~~~~+
 C([f'\arrowvert _{\textbf{x}\textbf{x}'=\textbf{c}\textbf{d}'}],[f'\arrowvert _{\textbf{x}\textbf{x}'
 =\textbf{c}\textbf{d}''}])(\tau ')\\
 &=C(\omega^{(d_1'-d_1'')g_{i_1}+\cdots+(d_p'-d_p'')g_{i_p}}[f\arrowvert _{\textbf{x}\textbf{x}'=\textbf{c}\textbf{d}''}],
 [f\arrowvert _{\textbf{x}\textbf{x}'=\textbf{c}\textbf{d}''}])(\tau ')\\&
 +C(\omega^{(d_1'-d_1'')g_{i_1}+\cdots+(d_p'-d_p'')g_{i_p}}[f'\arrowvert _{\textbf{x}\textbf{x}'=\textbf{c}\textbf{d}''}],
 [f'\arrowvert _{\textbf{x}\textbf{x}'=\textbf{c}\textbf{d}''}])(\tau ')\\
 &=\omega^{(d_1'-d_1'')g_{i_1}+\cdots+(d_p'-d_p'')g_{i_p}}\left( C([f\arrowvert _{\textbf{x}\textbf{x}'=\textbf{c}\textbf{d}''}],
 [f\arrowvert _{\textbf{x}\textbf{x}'=\textbf{c}\textbf{d}''}])(\tau ')\right. \\ &~~~~~~~~~~~~~~~~~~~~~~~~~~~~\left.
 +C([f'\arrowvert _{\textbf{x}\textbf{x}'=\textbf{c}\textbf{d}''}],[f'\arrowvert _{\textbf{x}\textbf{x}'=
 \textbf{c}\textbf{d}''}])(\tau ')\right)\\
& =\omega^{(d_1'-d_1'')g_{i_1}+\cdots+(d_p'-d_p'')g_{i_p}}\left( A([f\arrowvert _{\textbf{x}\textbf{x}'=\textbf{c}\textbf{d}''}])(\tau ')\right. \\&~~~~~~~~~~~~~~~~~~~~~~~~~~~~~~~~~ \left.
 +A([f'\arrowvert _{\textbf{x}\textbf{x}'=\textbf{c}\textbf{d}''}])(\tau ')\right).
 \end{split}
 \end{equation}
 Note that the  truncated vectors $[\psi(f\arrowvert _{\textbf{x}\textbf{x}'=\textbf{c}\textbf{d}''})]$ and $[\psi(f'\arrowvert _{\textbf{x}\textbf{x}'=\textbf{c}\textbf{d}''})]$ form a GCP. Therefore
 \begin{equation}
 \begin{split}
 A&([f\arrowvert _{\textbf{x}\textbf{x}'=\textbf{c}\textbf{d}''}])(\tau ')
 +A([f'\arrowvert _{\textbf{x}\textbf{x}'=\textbf{c}\textbf{d}''}])(\tau ')\\&=\begin{cases}
 2^{m-(k+p)+1}, & \tau'=0,\\
 0, & \text{otherwise}.
 \end{cases}
 \end{split}
 \end{equation}
 Substituting the value of auto-correlation sum into (\ref{exp1}), we have
 \begin{equation}
 \begin{split}
 C&([f\arrowvert _{\textbf{x}\textbf{x}'=\textbf{c}\textbf{d}'}],[f\arrowvert _{\textbf{x}\textbf{x}'=
 \textbf{c}\textbf{d}''}])(\tau ')\\&~~~~~~~~~~~~~~~~~~~+C([f'\arrowvert _{\textbf{x}\textbf{x}'
 =\textbf{c}\textbf{d}'}],[f'\arrowvert _{\textbf{x}\textbf{x}'=\textbf{c}\textbf{d}''}])(\tau ')\\&=
 \begin{cases}
 \omega^{(d_1'-d_1'')g_{i_1}+\cdots+(d_p'-d_p'')g_{i_p}}2^{m-(k+p)+1}, & \tau'=0,\\
 0, & \text{otherwise}.
 \end{cases}
 \end{split}
 \end{equation}
 The above cross-correlation sum is only nonzero at $\tau'=0$ i.e., when $\tau=u_1-u_2$, where $u_2$
 and $u_1$ are determined by $\textbf{x},\textbf{x}',\textbf{c},\textbf{d}'$ and $\textbf{d}''$, as follows.
 \begin{equation}\nonumber
 \begin{split}
 u_1=\displaystyle\sum_{\alpha=0}^{k-1}c_{\alpha}2^{j_{\alpha}}+d_1'2^{i_1}+\cdots+d_{p}'2^{i_{p}},\\
 u_2=\displaystyle\sum_{\alpha=0}^{k-1}c_{\alpha}2^{j_{\alpha}}+d_1''2^{i_1}+\cdots+d_{p}''2^{i_{p}},
 \end{split}
\end{equation}
where $c=(c_0,c_1,\cdots,c_{k-1})$. Hence $u_1-u_2=(d_1'-d_1'')2^{i_1}+\cdots+(d_{p}'-d_{p}'')2^{i_{p}}$.
Therefore the cross-correlation sum is  nonzero only at $\tau=(d_1'-d_1'')2^{i_1}+\cdots+(d_{p}'-d_{p}'')2^{i_{p}}$, where
the value is $\omega^{(d_1'-d_1'')g_{i_1}+\cdots+(d_p'-d_p'')g_{i_p}}2^{m-(k+p)+1}$, and thus \textit{Lemma} \ref{lemma6} is proved.
\end{IEEEproof}
 \end{lemma}
To illustrate \textit{Lemma \ref{lemma6}}, let us recall $R_{\tau_{i}}$ which is defined in (\ref{tau2}). Consider the GBFs $f$ and $f'$ of $5$ variables over $\mathbb{Z}_4$, as follows
\begin{equation}
\begin{split}
 f(x_0,x_1,x_2,x_3,x_4)&=2(x_2x_3+x_3x_1+x_0x_3+x_0x_1\\&~~~~~~+x_0x_2+x_0x_4),\\
 f'(x_0,x_1,x_2,x_3,x_4)&=2(x_2x_3+x_3x_1+x_0x_3+x_0x_1\\&~~~~~~+x_0x_2+x_0x_4+x_1).
 \end{split}
\end{equation}
Both $G(f\arrowvert_{x_0=c})$ and $G(f'\arrowvert_{x_0=c})$ $(c\in\mathbb{Z}_2)$ contain a path with $x_1$ as one of the end vertices and $x_4$ as
isolated vertex. Therefore $p=1$ and $i_1=4$. Hence the possible nonzero time-shifts are $\tau_0=(1-0)\cdot 2^4=16$, $\tau_1=(0-1)\cdot 2^4=-16$,
and the corresponding set of vectors are $R_{16}=\{(1,0)\}$, $R_{-16}=\{(0,1)\}$, respectively. By using \textit{Lemma} \ref{lemma6}, we show below that $R_{\tau_i}$'s are useful in the calculation of cross-correlation sum.
\begin{equation}
 \begin{split}
\nonumber
&\displaystyle\sum_{c'\neq c''}\left [C(f\arrowvert_{x_0x_4=cc'},f\arrowvert_{x_0x_4=cc''})\right. \\ &~~~~~~~~~~~~~~~~~~~~~~~~~~~~~~~\left.+C(f'\arrowvert_{x_0x_4=cc'},f'\arrowvert_{x_0x_4=cc''})\right ]\\
 &=\displaystyle\sum_{(c',c'')\in R_{16} }\left [C(f\arrowvert_{x_0x_4=cc'},f\arrowvert_{x_0x_4=cc''})\right. \\ &~~~~~~~~~~~~~~~~~~~~~~~~~~~~~~~~\left.+C(f'\arrowvert_{x_0x_4=cc'},f'\arrowvert_{x_0x_4=cc''})\right ]\\
 &~~+\displaystyle\sum_{(c',c'')\in R_{-16} }\left [ C(f\arrowvert_{x_0x_4=cc'},f\arrowvert_{x_0x_4=cc''})\right. \\ &~~~~~~~~~~~~~~~~~~~~~~~~~~~~~~~~\left.+C(f'\arrowvert_{x_0x_4=cc'},f'\arrowvert_{x_0x_4=cc''})\right ]\\
 &=C(f\arrowvert_{x_0x_4=c1},f\arrowvert_{x_0x_4=c0})+C(f'\arrowvert_{x_0x_4=c1},f'\arrowvert_{x_0x_4=c0})\\
&~~+C(f\arrowvert_{x_0x_4=c0},f\arrowvert_{x_0x_4=c1})+C(f'\arrowvert_{x_0x_4=c0},f'\arrowvert_{x_0x_4=c1})\\
&=\begin{cases}
   16, & \tau=\pm 16,\\
   0, & \textnormal{otherwise.}
  \end{cases}
 \end{split}
\end{equation}
 \begin{lemma}\cite{arthina}\label{lemma4}
Let $\textbf{d},\textbf{c}_1,\textbf{c}_2$ $\in \{0,1\}^k$. If $\textbf{c}_1\neq \textbf{c}_2$,
$\displaystyle\sum_{\textbf{d}}(-1)^{\textbf{d}\cdot(\textbf{c}_1+\textbf{c}_2)}=0$.
\end{lemma}
In the sequel, we provide the proof of \textit{Theorem} \ref{theorem1}.
 \begin{IEEEproof}
Let $f=Q+
 \displaystyle{\sum_{i=0}^{m-1}}g_ix_i+g'$ and the cross-correlation between $\psi(S_t)$, $\psi(S_{t'})$ can be written as
\begin{equation}\nonumber
\begin{split}
&C(\psi(S_t),\psi(S_{t'}))(\tau)\\
&=\displaystyle \sum_{\textbf{d}d}C\left(f+\frac{q}{2}\left((\textbf{d}+\textbf{b})\cdot\textbf{x}+\textbf{d}'\cdot\textbf{x}'
+dx_{\gamma}\right),\right. \\ &~~~~~~~~~~~~~~~~~~~~~\left.f+\frac{q}{2}\left((\textbf{d}+\textbf{b}')\cdot\textbf{x}+\textbf{d}''\cdot\textbf{x}'+
dx_{\gamma}\right) \right)(\tau)\qquad\qquad\qquad\qquad\qquad\\
&=S_1+S_2,\qquad\qquad\qquad\qquad\qquad\qquad\qquad\qquad\qquad\qquad\qquad\qquad\qquad\qquad\qquad\qquad\qquad\qquad\qquad\quad
\end{split}
\end{equation}
where
\begin{equation}\label{s1}
\begin{split}
S_1=\displaystyle \sum_{\textbf{d}d}\sum_{\textbf{c}_1\neq\textbf{c}_2}C\left(f+\frac{q}{2}\left((\textbf{d}+\textbf{b})\cdot\textbf{x}+\textbf{d}'\cdot\textbf{x}'
+dx_{\gamma}\right)\arrowvert_{\textbf{x}=\textbf{c}_1},\qquad\qquad\qquad\qquad\qquad\qquad\right.\\ \left. f+
\frac{q}{2}\left((\textbf{d}+\textbf{b}')\cdot\textbf{x}+\textbf{d}''\cdot\textbf{x}'+
dx_{\gamma}\right)\arrowvert_{\textbf{x}=\textbf{c}_2} \right)(\tau),\qquad\qquad\qquad\qquad\qquad\qquad
\end{split}
\end{equation}
and
\begin{equation}\label{s2}
\begin{split}
S_2=\displaystyle \sum_{\textbf{d}d}\sum_{\textbf{c}}C\left(f+\frac{q}{2}\left((\textbf{d}+\textbf{b})\cdot\textbf{x}+\textbf{d}'\cdot\textbf{x}'
+dx_{\gamma}\right)\arrowvert_{\textbf{x}=\textbf{c}},\qquad\qquad\qquad\qquad\qquad\qquad\right.\\ \left. f+
\frac{q}{2}\left((\textbf{d}+\textbf{b}')\cdot\textbf{x}+\textbf{d}''\cdot\textbf{x}'+dx_{\gamma}\right)\arrowvert_{\textbf{x}=\textbf{c}} \right)(\tau).\quad\qquad\qquad\qquad\qquad\qquad
\end{split}
\end{equation}
To find $S_1$, we start with \newline
\begin{equation}\label{pro}
\begin{split}
\sum_{\textbf{d}}C\left(f+\frac{q}{2}\left((\textbf{d}+\textbf{b})\cdot\textbf{x}+\textbf{d}'\cdot\textbf{x}'+
dx_{\gamma}\right)\arrowvert_{\textbf{x}=\textbf{c}_1},\qquad\qquad\qquad\qquad\qquad\qquad\right.\\ \left.  f+
\frac{q}{2}\left((\textbf{d}+\textbf{b}')\cdot\textbf{x}+\textbf{d}''\cdot\textbf{x}'+
dx_{\gamma}\right)\arrowvert_{\textbf{x}=\textbf{c}_2}\right)(\tau)\qquad\qquad\qquad\qquad\\
=(-1)^{\textbf{b}\cdot\textbf{c}_1+\textbf{b}'\cdot\textbf{c}_2}C\left(f+\frac{q}{2}\left(\textbf{d}'\cdot\textbf{x}'+
dx_{\gamma}\right)\arrowvert_{\textbf{x}=\textbf{c}_1},\qquad\qquad\qquad\qquad\qquad\qquad\right.\\ \left.  f+
\frac{q}{2}\left(\textbf{d}''\cdot\textbf{x}'+dx_{\gamma}\right)\arrowvert_{\textbf{x}=\textbf{c}_2}\right)(\tau)\sum_{\textbf{d}}(-1)^{\textbf{d}\cdot(\textbf{c}_1+\textbf{c}_2)}.\qquad\qquad\qquad\qquad
\end{split}
\end{equation}
By \textit{Lemma} \ref{lemma4} we have $\displaystyle\sum_{\textbf{d}}(-1)^{\textbf{d}\cdot(\textbf{c}_1+\textbf{c}_2)}=0$ for $\textbf{c}_1\neq \textbf{c}_2$,
therefore $S_1$ vanishes for all values of $\tau$.
Similarly, to simplify $S_2$ we start with
\begin{equation}\label{find S_1}
\begin{split}
&\sum_{\textbf{d}}C\left(f+\frac{q}{2}\left((\textbf{d}+\textbf{b})\cdot\textbf{x}+\textbf{d}'\cdot\textbf{x}'+
dx_{\gamma}\right)\arrowvert_{\textbf{x}=\textbf{c}},\right. \\ &~~~~~~~~~~~~~\left.f+
\frac{q}{2}\left((\textbf{d}+\textbf{b}')\cdot\textbf{x}+\textbf{d}''\cdot\textbf{x}'+dx_{\gamma}\right)\arrowvert_{\textbf{x}=\textbf{c}}\right)(\tau)\\
&=(-1)^{\textbf{b}\cdot\textbf{c}+\textbf{b}'\cdot\textbf{c}}C\left(f+\frac{q}{2}\left(\textbf{d}'\cdot\textbf{x}'+
dx_{\gamma}\right)\arrowvert_{\textbf{x}=\textbf{c}},\right. \\&~~~~~~~~~~~~~\left.
f+\frac{q}{2}\left(\textbf{d}''\cdot\textbf{x}'+dx_{\gamma}\right)\arrowvert_{\textbf{x}=\textbf{c}}\right)(\tau)\sum_{\textbf{d}}
(-1)^{\textbf{d}\cdot(\textbf{c}+\textbf{c})}\\
&=(-1)^{\textbf{b}\cdot\textbf{c}+\textbf{b}'\cdot\textbf{c}} 2^k C\left(f+\frac{q}{2}\left(\textbf{d}'\cdot\textbf{x}'+
dx_{\gamma}\right)\arrowvert_{\textbf{x}=\textbf{c}},\right. \\ &~~~~~~~~~~~~~~~~~~~~~\qquad~~\left.
f+\frac{q}{2}\left(\textbf{d}''\cdot\textbf{x}'+dx_{\gamma}\right)\arrowvert_{\textbf{x}=\textbf{c}}\right)(\tau)\\
&=(-1)^{\textbf{b}\cdot\textbf{c}+\textbf{b}'\cdot\textbf{c}} 2^k\sum_{\textbf{c}'\textbf{c}''} C\left(f+\frac{q}{2}\left(\textbf{d}'\cdot\textbf{x}'
+dx_{\gamma}\right)\arrowvert_{\textbf{x}\textbf{x}'=\textbf{c}\textbf{c}'},\right. \\ &~~~~~~~~~~~~~~~~~~~~~~~~~\left.
f+\frac{q}{2}\left(\textbf{d}''\cdot\textbf{x}'+dx_{\gamma}\right)\arrowvert_{\textbf{x}\textbf{x}'=\textbf{c}\textbf{c}''}\right)(\tau)\\
&=(-1)^{\textbf{b}\cdot\textbf{c}+\textbf{b}'\cdot\textbf{c}} 2^k\sum_{\textbf{c}'\textbf{c}''}(-1)^{\textbf{d}'\cdot\textbf{c}'
+\textbf{d}''\cdot\textbf{c}''} C\left(f+\frac{q}{2}dx_{\gamma}\arrowvert_{\textbf{x}\textbf{x}'=\textbf{c}\textbf{c}'},\right. \\&~~~~~~~~~~~~~~~~~~~~~~~~~~~~~~~~~~~~~~\left.
f+\frac{q}{2}dx_{\gamma}\arrowvert_{\textbf{x}\textbf{x}'=\textbf{c}\textbf{c}''}\right)(\tau)\\
&\!=\!(-1)^{\textbf{b}\cdot\textbf{c}+\textbf{b}'\cdot\textbf{c}} 2^k\!\!
\left(\sum_{\textbf{c}'=\textbf{c}''}(-1)^{\textbf{d}'\cdot\textbf{c}'+\textbf{d}''\cdot\textbf{c}''}\right. C\left(f\!\!+\!\!\frac{q}{2}
dx_{\gamma}\arrowvert_{\textbf{x}\textbf{x}'=\textbf{c}\textbf{c}'},\left.\right.\right.\\&~~~~~~~~~~~~~~~~~~~~~~~~~~~~~~~~~~~\left.\left. f+
\frac{q}{2}dx_{\gamma}\arrowvert_{\textbf{x}\textbf{x}'=\textbf{c}\textbf{c}''}\right)(\tau)\right.\\ &~~~~+
\left.\sum_{\textbf{c}'\neq \textbf{c}''}(-1)^{\textbf{d}'\cdot\textbf{c}'+\textbf{d}''\cdot\textbf{c}''}\right. C\left(f+
\frac{q}{2}dx_{\gamma}\arrowvert_{\textbf{x}\textbf{x}'=\textbf{c}\textbf{c}'}, \left.\right.\right.\\&~~~~~~~~~~~~~~~~~~~~~~~~~~~~~~~~~~~~\left.\left.  f
+\frac{q}{2}dx_{\gamma}\arrowvert_{\textbf{x}\textbf{x}'=\textbf{c}\textbf{c}''}\right)(\tau) \right)\\
&\!\!=\!\!(-1)^{\textbf{b}\cdot\textbf{c}+\textbf{b}'\cdot\textbf{c}} 2^k\!\!\left(\sum_{\textbf{c}'=\textbf{c}''}\!\!(-1)^{(\textbf{d}'\cdot+\textbf{d}'')
\cdot\textbf{c}'}\right.\!\!\!\! A\left(f\!\!+\!\!\frac{q}{2}dx_{\gamma}\arrowvert_{\textbf{x}\textbf{x}'=\textbf{c}\textbf{c}'}\right)\!\!(\tau)\\
&~~~~+\left.\sum_{\textbf{c}'\neq \textbf{c}''}(-1)^{\textbf{d}'\cdot\textbf{c}'+\textbf{d}''\cdot\textbf{c}''}\right. C\left(f+
\frac{q}{2}dx_{\gamma}\arrowvert_{\textbf{x}\textbf{x}'=\textbf{c}\textbf{c}'},\left.\right.\right.\\&~~~~~~~~~~~~~~~~~~~~~~~~~~~~~~~~~\left.\left.  f+
\frac{q}{2}dx_{\gamma}\arrowvert_{\textbf{x}\textbf{x}'=\textbf{c}\textbf{c}''}\right)(\tau) \right)\\
&=(-1)^{\textbf{b}\cdot\textbf{c}+\textbf{b}'\cdot\textbf{c}} 2^k(L_1+L_2),\quad\qquad\qquad\qquad\qquad\qquad\qquad\qquad\qquad\qquad\quad
\end{split}
\end{equation}
where
\begin{equation}\label{pal1}
L_1=\sum_{\textbf{c}'=\textbf{c}''}(-1)^{(\textbf{d}'+\textbf{d}'')\cdot\textbf{c}'} A\left(f
+\frac{q}{2}dx_{\gamma}\arrowvert_{\textbf{x}\textbf{x}'=\textbf{c}\textbf{c}'}\right)(\tau),\qquad\qquad\qquad
\end{equation}
and
\begin{equation}
\begin{split}
L_2=\sum_{\textbf{c}'\neq \textbf{c}''}(-1)^{\textbf{d}'\cdot\textbf{c}'+\textbf{d}''\cdot\textbf{c}''} C\left(f
+\frac{q}{2}dx_{\gamma}\arrowvert_{\textbf{x}\textbf{x}'=\textbf{c}\textbf{c}'},\right. \qquad\qquad\qquad\\ \left.  f
+\frac{q}{2}dx_{\gamma}\arrowvert_{\textbf{x}\textbf{x}'=\textbf{c}\textbf{c}''}\right)(\tau).\qquad\qquad
\end{split}
\end{equation}
Since $G(f\arrowvert_{\textbf{xx}'=\textbf{cc}'})$ is a path over $m-k-p$ ( $p$ is the number of isolated vertices) vertices,\\
\begin{equation}\label{pal2}
\displaystyle\sum_{d}A\left(f+\frac{q}{2}dx_{\gamma}\arrowvert_{\textbf{x}\textbf{x}'=\textbf{c}\textbf{c}'}\right)(\tau)=
\begin{cases}
2^{m-(k+p)+1}, & \tau=0,\\
0, & \text{otherwise}.
\end{cases}
\end{equation}
Now from (\ref{pal1}) and (\ref{pal2}) we have
\begin{equation}\label{pal11}
\displaystyle\sum_{d}L_1=
\begin{cases}
2^{m-(k+p)+1}\displaystyle\sum_{\textbf{c}'=\textbf{c}''}(-1)^{(\textbf{d}'+\textbf{d}'')\cdot\textbf{c}'}, & \tau=0,\\
0, & \text{otherwise}.
\end{cases}
\end{equation}
Therefore, \begin{equation}\label{pal3}
\displaystyle\sum_{d}L_1=
\begin{cases}
2^{m-k+1}, & \tau=0, \textbf{d}'=\textbf{d}'', \\
0, & \tau=0, \textbf{d}'\neq\textbf{d}'', \\
0, & \text{otherwise}.\qquad\qquad\qquad\qquad\qquad
\end{cases}
\end{equation}
To find simplified value of $L_2$, we start with
\begin{equation}\nonumber
\{(\textbf{c}',\textbf{c}''): \textbf{c}',\textbf{c}''\in \mathbb{Z}_q \quad\text{and} \quad \textbf{c}'\neq\textbf{c}''\}=
\cup_{i=1}^r R_{\tau_i}.\qquad\qquad\qquad\qquad
\end{equation}
Therefore
\begin{equation}\label{final L_2}
\begin{split}
L_2
&=\sum_{\textbf{c}'\neq \textbf{c}''}(-1)^{\textbf{d}'\cdot\textbf{c}'+\textbf{d}''\cdot\textbf{c}''} C\left(f
+\frac{q}{2}dx_{\gamma}\arrowvert_{\textbf{x}\textbf{x}'=\textbf{c}\textbf{c}'},\right.\\ &~~~~~~~~~~~~~~~~~~~~~~~~~~~~~~~~~~~\left. f
+\frac{q}{2}dx_{\gamma}\arrowvert_{\textbf{x}\textbf{x}'=\textbf{c}\textbf{c}''}\right)(\tau)\\
&=\sum_{(\textbf{c}',\textbf{c}'')\in \cup_{i=1}^rR_{\tau_i}}(-1)^{\textbf{d}'\cdot\textbf{c}'+\textbf{d}''\cdot\textbf{c}''} C\left(f+
\frac{q}{2}dx_{\gamma}\arrowvert_{\textbf{x}\textbf{x}'=\textbf{c}\textbf{c}'},\right.\\&~~~~~~~~~~~~~~~~~~~~~~~~~~~~~~~~~~~\left.f
+\frac{q}{2}dx_{\gamma}\arrowvert_{\textbf{x}\textbf{x}'=\textbf{c}\textbf{c}''}\right)(\tau)\\
&=\sum_{i=1}^{r}\sum_{(\textbf{c}',\textbf{c}'')\in R_{\tau_i}}(-1)^{\textbf{d}'\cdot\textbf{c}'+\textbf{d}''\cdot\textbf{c}''} C\left(f
+\frac{q}{2}dx_{\gamma}\arrowvert_{\textbf{x}\textbf{x}'=\textbf{c}\textbf{c}'},\right.\\&~~~~~~~~~~~~~~~~~~~~~~~~~~~~~~~~~~~~~\left. f
+\frac{q}{2}dx_{\gamma}\arrowvert_{\textbf{x}\textbf{x}'=\textbf{c}\textbf{c}''}\right)(\tau).
\end{split}
\end{equation}
Since the vertices $x_{i_1}, x_{i_2}, \cdots x_{i_{p}}$ are isolated by the deletion operations, in the function $f$ the only quadratic terms
involving variables $x_{i_\beta}$'s are those with the variables of the deleted vertices. Thus the only quadratic  terms in
$x_{i_\beta}$'s in $f$ can be expressed as follows.
\begin{equation}
\sum_{\alpha=0}^{k-1}\sum_{\beta=1}^{p}e'_{\alpha,\beta}x_{j_\alpha}x_{i_\beta},
\end{equation}
where $e'_{\alpha,\beta}$ are the weights of the edges between
the deleted vertices and the isolated vertices. Now the term $\left(f
+\frac{q}{2}dx_{\gamma}\right)\arrowvert_{\textbf{x}\textbf{x}'=\textbf{c}\textbf{c}'}$ can be written as
\begin{equation}\nonumber
\begin{split}
&\left(f
+\frac{q}{2}dx_{\gamma}\right)\arrowvert_{\textbf{x}\textbf{x}'=\textbf{c}\textbf{c}'}\\
&=\left(f'+\displaystyle\sum_{\alpha=0}^{k-1}\left(\sum_{\beta=1}^{p}e'_{\alpha,\beta}x_{j_\alpha}x_{i_\beta}\right)
+\frac{q}{2}dx_{\gamma}\right)\arrowvert_{\textbf{x}\textbf{x}'=\textbf{c}\textbf{c}'}\\
&=\left(f'+\displaystyle\sum_{\alpha=0}^{k-1}\left(\sum_{\beta=1}^{p}e'_{\alpha,\beta}c_{\alpha}c_{\beta}'\right)
+\frac{q}{2}dx_{\gamma}\right)\arrowvert_{\textbf{x}\textbf{x}'=\textbf{c}\textbf{c}'}\\
&\!=\!\omega^{g_{\textbf{c}\textbf{c}'}}\left(f'
\!\!+\!\!\frac{q}{2}dx_{\gamma}\right)\arrowvert_{\textbf{x}\textbf{x}'=\textbf{c}\textbf{c}'}(\text{where}~g_{\textbf{c}\textbf{c}'}\triangleq\sum_{\alpha=0}^{k-1}\sum_{\beta=1}^{p}e'_{\alpha,\beta}
c_{\alpha}c_{\beta}').
\end{split}
\end{equation}
Similarly
\begin{equation}
\begin{split}
 &\left(f
+\frac{q}{2}dx_{\gamma}\right)\arrowvert_{\textbf{x}\textbf{x}''=\textbf{c}\textbf{c}''}\\
&\!=\!\omega^{g_{\textbf{c}\textbf{c}''}}\left(f'
\!\!+\!\!\frac{q}{2}dx_{\gamma}\right)\arrowvert_{\textbf{x}\textbf{x}''=\textbf{c}\textbf{c}''}(\text{where}~g_{\textbf{c}\textbf{c}''}\!\!\triangleq\!\!\sum_{\alpha=0}^{k-1}\sum_{\beta=1}^{p}e'_{\alpha,\beta}
c_{\alpha}c_{\beta}'').
\end{split}
\end{equation}
Therefore, the term $C\left(f
\!\!+\!\!\frac{q}{2}dx_{\gamma}\arrowvert_{\textbf{x}\textbf{x}'=\textbf{c}\textbf{c}'}, f
\!\!+\!\!\frac{q}{2}dx_{\gamma}\arrowvert_{\textbf{x}\textbf{x}'=\textbf{c}\textbf{c}''}\right)(\tau)$ can be simplified to
\begin{equation}
\begin{split}
 &C\left(f
+\frac{q}{2}dx_{\gamma}\arrowvert_{\textbf{x}\textbf{x}'=\textbf{c}\textbf{c}'},f
+\frac{q}{2}dx_{\gamma}\arrowvert_{\textbf{x}\textbf{x}'=\textbf{c}\textbf{c}''}\right)(\tau)\\
&=\omega^{g_{\textbf{c}\textbf{c}'}-g_{\textbf{c}\textbf{c}''}}C\left(f'
+\frac{q}{2}dx_{\gamma}\arrowvert_{\textbf{x}\textbf{x}'=\textbf{c}\textbf{c}'}, f'
+\frac{q}{2}dx_{\gamma}\arrowvert_{\textbf{x}\textbf{x}'=\textbf{c}\textbf{c}''}\right)(\tau).
\end{split}
\end{equation}
By \textit{Lemma} \ref{lemma6}, we have
 \begin{equation}\label{finallemma}
 \begin{split}
  &\displaystyle\sum_{d}C\left(f'+\frac{q}{2}dx_{\gamma}\arrowvert_{\textbf{x}\textbf{x}'=\textbf{c}\textbf{c}'},
    f'+\frac{q}{2}dx_{\gamma}\arrowvert_{\textbf{x}\textbf{x}'=\textbf{c}\textbf{c}''}\right)(\tau) \\
&=\begin{cases}
2^{m-(k+p)+1}\omega^{(c_1'-c_1'')g_{i_1}+\cdots+(c_p'-c_p'')g_{i_p}},\\~~~~~~~~~~~~~~~~ \tau=(c_1'-c_1'')2^{i_1}+\cdots+(c_p'-c_p'')2^{i_p},\\
0, ~~~~~~~~~~~~~~\text{otherwise}.
\end{cases}\\
&=\begin{cases}
2^{m-(k+p)+1}\omega^{(\textbf{c}'-\textbf{c}'')\cdot \varGamma}, & \tau=T_{(\textbf{c}'-\textbf{c}'')},\\
0, & \text{otherwise}.
\end{cases}
\end{split}
\end{equation}
 Now from (\ref{final L_2}) and (\ref{finallemma}) we have,
 \begin{equation} \label{final1 L_2}
\begin{split}
&\displaystyle\sum_{d}L_2\\
&=\sum_{i=1}^{r}\sum_{(\textbf{c}',\textbf{c}'')\in R_{\tau_i}}(-1)^{\textbf{d}'\cdot\textbf{c}'+\textbf{d}''\cdot\textbf{c}''}
\displaystyle\sum_{d}C\left(f+\frac{q}{2}dx_{\gamma}\arrowvert_{\textbf{x}\textbf{x}'=\textbf{c}\textbf{c}'},\right.\\&\left.~~~~~~~~~~~~~~~~~~~~~~~~~~~~~~~~~~~~~~~~~~ f
+\frac{q}{2}dx_{\gamma}\arrowvert_{\textbf{x}\textbf{x}'=\textbf{c}\textbf{c}''}\right)(\tau)\\
&=\begin{cases}
2^{m-(k+p)+1}\\ \times\displaystyle\sum_{(\textbf{c}',\textbf{c}'')\in R_{\tau}}\omega^{g_{\textbf{c}\textbf{c}'}-g_{\textbf{c}\textbf{c}''}}(-1)^{\textbf{d}'\cdot\textbf{c}'+\textbf{d}''\cdot\textbf{c}''}\omega^{(\textbf{c}'-\textbf{c}'')\cdot \varGamma}, \\
~~~~~~~~~~~~~~~~~~~~~~~~~~~~~~~~~~~~\tau=\tau_i, i=1,2,\cdots,r,\\
0, ~~~~~~~~~~~~~~~~~~~~~~~~~~~~~~~~~~\textnormal{otherwise}.
\end{cases}
\end{split}
\end{equation}
For $\textbf{d}'=\textbf{d}''$, from (\ref{pal3}) and (\ref{final1 L_2}) we have
\begin{equation} \label{final2 L_2}
\begin{split}
&\displaystyle\sum_{d}(L_1+L_2)\\
&=\begin{cases}
2^{m-k+1},~~~~~~~~~~~~~~~~~~~~~~ \tau=0, \\
2^{m-(k+p)+1}\\\times\displaystyle\sum_{(\textbf{c}',\textbf{c}'')\in R_{\tau}}\omega^{g_{\textbf{c}\textbf{c}'}-g_{\textbf{c}\textbf{c}''}}(-1)^{\textbf{d}'\cdot(\textbf{c}'+\textbf{c}'')}\omega^{(\textbf{c}'-\textbf{c}'')\cdot \varGamma}, \\
~~~~~~~~~~~~~~~~~~~~~~~~~~~~~~~~~~\tau=\tau_i, i=1,2,\cdots,r,\\
0,~~~~~~~~~~~~~~~~~~~~~~~~~~~~~~~~ \text{otherwise}.
\end{cases}
\end{split}
\end{equation}
If $\textbf{d}'\neq\textbf{d}''$, from (\ref{pal3}) and (\ref{final1 L_2}) we have
\begin{equation}
\begin{split}
&\displaystyle\sum_{d}(L_1+L_2)\\
&=\begin{cases}
2^{m-(k+p)+1}\\ \times\displaystyle\sum_{(\textbf{c}',\textbf{c}'')\in R_{\tau}}\omega^{g_{\textbf{c}\textbf{c}'}-g_{\textbf{c}\textbf{c}''}}(-1)^{\textbf{d}'\cdot\textbf{c}'+\textbf{d}''\cdot\textbf{c}''}\omega^{(\textbf{c}'-\textbf{c}'')\cdot \varGamma},\\
~~~~~~~~~~~~~~~~~~~~~~~~~~~~~~~~~~~\tau=\tau_i, i=1,2,\cdots,r,\\
0,~~~~~~~~~~~~~~~~~~~~~~~~~~~~~~~~~   \text{otherwise}.
\end{cases}
\end{split}
\end{equation}
For $\textbf{d}'=\textbf{d}''$, from (\ref{find S_1}), (\ref{final2 L_2}) we have
\begin{equation}\label{sam1}
\begin{split}
&\sum_{\textbf{d},d}C\left(f+\frac{q}{2}\left((\textbf{d+b})\cdot\textbf{x}+\textbf{d}'\cdot\textbf{x}'+
dx_{\gamma}\right)\arrowvert_{\textbf{x}=\textbf{c}},\right.\\&~~~~~~~~~~~~~~~\left.f
+\frac{q}{2}\left((\textbf{d}+\textbf{b}')\cdot\textbf{x}+\textbf{d}''\cdot\textbf{x}'+dx_{\gamma}\right)\arrowvert_{\textbf{x}=\textbf{c}}\right)(\tau)\\
&=(-1)^{\textbf{b}\cdot\textbf{c}+\textbf{b}'\cdot\textbf{c}}2^k \displaystyle\sum_{d}(L_1+L_2)\\
&=\begin{cases}
2^{m+1}(-1)^{\textbf{b}\cdot\textbf{c}+\textbf{b}'\cdot\textbf{c}}, ~~~~~~~~~~~\tau=0, \\
2^{m-p+1}(-1)^{\textbf{b}\cdot\textbf{c}+\textbf{b}'\cdot\textbf{c}}\\ \times\displaystyle\sum_{(\textbf{c}',\textbf{c}'')\in R_{\tau}}\omega^{g_{\textbf{c}\textbf{c}'}-g_{\textbf{c}\textbf{c}''}}(-1)^{\textbf{d}'\cdot(\textbf{c}'+\textbf{c}'')}
\omega^{(\textbf{c}'-\textbf{c}'')\cdot \varGamma},\\~~~~~~~~~~~~~~~~~~~~~~~~~~~~~~~~~ \tau=\tau_i, i=1,2,\cdots,r,\\
0, ~~~~~~~~~~~~~~~~~~~~~~~~~~~~~~~\text{otherwise}.
\end{cases}
\end{split}
\end{equation}
For $\textbf{d}'\neq\textbf{d}''$, from (\ref{find S_1}) and (\ref{final1 L_2}) we have
\begin{equation}\label{sam2}
\begin{split}
&\sum_{\textbf{d},d}C\left(f+\frac{q}{2}\left((\textbf{d+b})\cdot\textbf{x}+\textbf{d}'\cdot\textbf{x}'+
dx_{\gamma}\right)\arrowvert_{\textbf{x}=\textbf{c}},\right.\\&~~~~~~~~~~\left.f
+\frac{q}{2}\left((\textbf{d}+\textbf{b}')\cdot\textbf{x}+\textbf{d}''\cdot\textbf{x}'+dx_{\gamma}\right)\arrowvert_{\textbf{x}=\textbf{c}}\right)(\tau)\\
&=(-1)^{\textbf{b}\cdot\textbf{c}+\textbf{b}'\cdot\textbf{c}}2^k \displaystyle\sum_{d}(L_1+L_2)\\
&=\begin{cases}
2^{m-p+1}(-1)^{\textbf{b}\cdot\textbf{c}+\textbf{b}'\cdot\textbf{c}}\\\times\displaystyle\sum_{(\textbf{c}',\textbf{c}'')\in R_{\tau}}\omega^{g_{\textbf{c}\textbf{c}'}-g_{\textbf{c}\textbf{c}''}}
(-1)^{\textbf{d}'\cdot\textbf{c}'+\textbf{d}''\cdot\textbf{c}''}\omega^{(\textbf{c}'-\textbf{c}'')\cdot \varGamma}, \\~~~~~~~~~~~~~~~~~~~~~~~~~~~~~~~~~~~ \tau=\tau_i, i=1,2,\cdots,r,\\
0,~~~~~~~~~~~~~~~~~~~~~~~~~~~~~~~~~\text{otherwise}.
\end{cases}\qquad\qquad
\end{split}
\end{equation}
For $\textbf{d}'=\textbf{d}''$, $\textbf{b}=\textbf{b}'$, from (\ref{s1}), (\ref{sam1}) and using \textit{Lemma} \ref{lemma6} we have
\begin{equation}\label{sam4}
\begin{split}
S_2&=\displaystyle \sum_{\textbf{d}d}\sum_{\textbf{c}}C\left(f\!+\!\frac{q}{2}\left((\textbf{d+b})\!\cdot\!\textbf{x}+\textbf{d}'\!\cdot\!\textbf{x}'
+dx_{\gamma}\right)\arrowvert_{\textbf{x}=\textbf{c}},\right.\\&~~~~~~~~~~~~~~\left.f
+\frac{q}{2}\left((\textbf{d}+\textbf{b}')\!\cdot\!\textbf{x}+\textbf{d}''\!\cdot\!\textbf{x}'+dx_{\gamma}\right)\arrowvert_{\textbf{x}=\textbf{c}} \right)(\tau)\\
&=\begin{cases}
2^{m+1}\displaystyle\sum_{\textbf{c}}(-1)^{\textbf{b}\cdot\textbf{c}+\textbf{b}'\cdot\textbf{c}},~~~~~~~\tau=0, \\
2^{m-p+1}\displaystyle\sum_{(\textbf{c}',\textbf{c}'')\in R_{\tau}}
(-1)^{\textbf{d}'\cdot(\textbf{c}'+\textbf{c}'')}\omega^{(\textbf{c}'-\textbf{c}'')\cdot \varGamma}\\ \times
\displaystyle\sum_{\textbf{c}}\omega^{g_{\textbf{c}\textbf{c}'}-g_{\textbf{c}\textbf{c}''}}(-1)^{\textbf{b}\cdot\textbf{c}+\textbf{b}'\cdot\textbf{c}}, \tau=\tau_i, i=1,2,\cdots,r,\\
0,~~~~~~~~~~~~~~~~~~~~~~~~~~~~~~~~\text{otherwise}.
\end{cases}
\end{split}
\end{equation}
\begin{equation}\nonumber
\begin{split}
=\begin{cases}
2^{m+k+1},~~~~~~~~~~~~~~~~~~~~~\tau=0, \\
2^{m-p+1}\displaystyle\sum_{(\textbf{c}',\textbf{c}'')\in R_{\tau}}
(-1)^{\textbf{d}'\cdot(\textbf{c}'+\textbf{c}'')}\omega^{(\textbf{c}'-\textbf{c}'')\cdot \varGamma}\\ \times\displaystyle\sum_{\textbf{c}}
\omega^{g_{\textbf{c}\textbf{c}'}-g_{\textbf{c}\textbf{c}''}},~~~~~~~~~~~~\tau=\tau_i, i=1,2,\cdots,r,\\
0,~~~~~~~~~~~~~~~~~~~~~~~~~~~~~\text{otherwise}.
\end{cases}\qquad\qquad\quad
\end{split}
\end{equation}
For $\textbf{d}'\neq\textbf{d}''$, $\textbf{b}=\textbf{b}'$, from (\ref{s1}), (\ref{sam2}) and using \textit{Lemma} \ref{lemma6} we have
\begin{equation}\label{sam5}
\begin{split}
S_2&=\displaystyle \sum_{\textbf{d}d}\sum_{\textbf{c}}C\left(f+\frac{q}{2}\left((\textbf{d+b})\!\cdot\!\textbf{x}+\textbf{d}'\!\cdot\!\textbf{x}'+dx_{\gamma}\right)\arrowvert_{\textbf{x}=\textbf{c}}, \right. \\ &~~~~~~~~~~~~~\left. f
+\frac{q}{2}\left((\textbf{d}+\textbf{b}')\!\cdot\!\textbf{x}+\textbf{d}''\!\cdot\!\textbf{x}'+dx_{\gamma}\right)\arrowvert_{\textbf{x}=\textbf{c}} \right)(\tau)\\
&=\begin{cases}
2^{m-p+1}\displaystyle\sum_{(\textbf{c}',\textbf{c}'')\in R_{\tau}}(-1)^{\textbf{d}'\cdot\textbf{c}'+\textbf{d}''\cdot\textbf{c}''}
\omega^{(\textbf{c}'-\textbf{c}'')\cdot \varGamma}\\ \times
\displaystyle\sum_{\textbf{c}}\omega^{g_{\textbf{c}\textbf{c}'}-g_{\textbf{c}\textbf{c}''}}(-1)^{\textbf{b}\cdot\textbf{c}+\textbf{b}'\cdot\textbf{c}},
~\tau=\tau_i, i=1,2,\cdots,r,\\
0,~~~~~~~~~~~~~~~~~~~~~~~~~~~~~~~~ \text{otherwise}.
\end{cases}\\
&=\begin{cases}
2^{m-p+1}\displaystyle\sum_{(\textbf{c}',\textbf{c}'')\in R_{\tau}}(-1)^{\textbf{d}'\cdot\textbf{c}'+\textbf{d}''\cdot\textbf{c}''}
\omega^{(\textbf{c}'-\textbf{c}'')\cdot \varGamma}\\ \times\sum_{\textbf{c}}\omega^{g_{\textbf{c}\textbf{c}'}-g_{\textbf{c}\textbf{c}''}}, ~\tau=\tau_i, i=1,2,\cdots,r,\\
0,~~~~~~~~~~~~~~~~~~ \text{otherwise}.
\end{cases}
\end{split}
\end{equation}
For $\textbf{d}'=\textbf{d}''$, $\textbf{b}\neq\textbf{b}'$, from (\ref{sam4}) and \textit{Lemma} \ref{lemma6} we have
\begin{equation}\label{sam6}
\begin{split}
S_2&=\displaystyle \sum_{\textbf{d}d}\sum_{\textbf{c}}C\left(f+\frac{q}{2}\left((\textbf{d+b})\!\cdot\!\textbf{x}+\textbf{d}'\!\cdot\!\textbf{x}'
+dx_{\gamma}\right)\arrowvert_{\textbf{x}=\textbf{c}},\right. \\&~~~~~~~ \left.f+\frac{q}{2}\left((\textbf{d}+
\textbf{b}')\!\cdot\!\textbf{x}+\textbf{d}''\cdot\textbf{x}'+dx_{\gamma}\right)\arrowvert_{\textbf{x}=\textbf{c}} \right)(\tau)\\
&=\begin{cases}
2^{m+1}\displaystyle\sum_{\textbf{c}}(-1)^{\textbf{b}\cdot\textbf{c}+\textbf{b}'\cdot\textbf{c}},~~~~\tau=0, \\
2^{m-p+1}\displaystyle\sum_{(\textbf{c}',\textbf{c}'')\in R_{\tau}}(-1)^{\textbf{d}'\cdot(\textbf{c}'+\textbf{c}'')}
\omega^{(\textbf{c}'-\textbf{c}'')\cdot \varGamma}\\
\displaystyle\sum_{\textbf{c}}\omega^{g_{\textbf{c}\textbf{c}'}-g_{\textbf{c}\textbf{c}''}}(-1)^{\textbf{b}\cdot\textbf{c}+\textbf{b}'\cdot\textbf{c}},
\tau=\tau_i, i=1,2,\cdots,r,\\
0, ~~~~~~~~~~~~~~~~~~~~~~~~~~~~~\text{otherwise}.
\end{cases}\\
&=\begin{cases}
2^{m-p+1}\displaystyle\sum_{(\textbf{c}',\textbf{c}'')\in R_{\tau}}(-1)^{\textbf{d}'.(\textbf{c}'+\textbf{c}'')}
\omega^{(\textbf{c}'-\textbf{c}'')\cdot \varGamma}\\ \times\displaystyle\sum_{\textbf{c}}\omega^{g_{\textbf{c}\textbf{c}'}-g_{\textbf{c}\textbf{c}''}}(-1)^{\textbf{b}\cdot\textbf{c}+\textbf{b}'\cdot\textbf{c}},
~\tau\!=\!\tau_i, i\!=\!1,2,\cdots,r,\\
0, ~~~~~~~~~~~~~~~~~~~~~~~~~~~~~~~~\text{otherwise}.
\end{cases}\end{split}
\end{equation}
For $\textbf{d}'\neq\textbf{d}''$, $\textbf{b}\neq\textbf{b}'$, from (\ref{sam5}) and using \textit{Lemma} \ref{lemma6} we have
\begin{equation}\label{sam7}
\begin{split}
S_2&=\displaystyle \sum_{\textbf{d}d}\sum_{\textbf{c}}C\left(f+\frac{q}{2}\left((\textbf{d+b})\!\cdot\!\textbf{x}+\textbf{d}'\!\cdot\!\textbf{x}'+
dx_{\gamma}\right)\arrowvert_{\textbf{x}=\textbf{c}},\right. \\ &~~~~~~~~~~~~\left.f+
\frac{q}{2}\left((\textbf{d}+\textbf{b}')\!\cdot\!\textbf{x}+\textbf{d}''\cdot\textbf{x}'+dx_{\gamma}\right)\arrowvert_{\textbf{x}=\textbf{c}} \right)(\tau)\\
&=\begin{cases}
2^{m-p+1}\displaystyle\sum_{(\textbf{c}',\textbf{c}'')\in R_{\tau}}(-1)^{\textbf{d}'\cdot\textbf{c}'+\textbf{d}''\cdot\textbf{c}''}
\omega^{(\textbf{c}'-\textbf{c}'')\cdot \varGamma}\\ \times\displaystyle\sum_{\textbf{c}}\omega^{g_{\textbf{c}\textbf{c}'}-g_{\textbf{c}\textbf{c}''}}(-1)^{\textbf{b}\cdot\textbf{c}+\textbf{b}'\cdot\textbf{c}},
\tau=\tau_i, i=1,2,\cdots,r,\\
0, ~~~~~~~~~~~~~~~~~~~~~~~~~~~~~~~\text{otherwise}.
\end{cases}
\end{split}
\end{equation}
The result in (\ref{sam4}) proves the hypothesis 1 given in (\ref{auto1}).\\
(\ref{sam5}), (\ref{sam6}) and (\ref{sam7}) prove the hypothesis 2 given in (\ref{cross1}).\\
\end{IEEEproof}

 \section{Proof of \textnormal{\textit{Theorem} \ref{theorem2}}}

 \begin{IEEEproof}
Let $t'=\displaystyle\sum_{\alpha=0}^{k-1}b'_{\alpha}2^{\alpha}+\sum_{\alpha=k}^{k+p-1}d'_{\alpha}2^{\alpha}$ and
 $t''=\displaystyle\sum_{\alpha=0}^{k-1}b''_{\alpha}2^{\alpha}+\sum_{\alpha=k}^{k+p-1}d''_{\alpha}2^{\alpha}$ $(b'_{\alpha},b''_{\alpha},d'_{\alpha},d''_{\alpha}\in \mathbb{Z}_2\hspace*{0.1cm} \text{and}\hspace*{0.1cm} 0\leq t',t''\leq 2^{k+p}-1)$.
Since the  labels of the isolated vertices are $m-p,m-p+1,\cdots,m-1$, we consider $\textbf{x}'=(x_{m-p},x_{m-p+1},$ $\cdots,x_{m-1})$ and
$\varGamma=(g_{m-p},g_{m-p+1},\cdots,g_{m-1})$.\\
 To prove \textit{Theorem} \ref{theorem2}, we only need to show \begin{equation}\nonumber
C(\psi(S_t'),\psi^*(\bar{S}_{t''}))=0, \quad\left |\tau\right |\leq 2^{m-p}.
\end{equation}
Let us start with
\begin{equation}\label{sab3}
 \begin{split}
&C(\psi(S_{t'}),\psi^*(\bar{S}_{t''}))(\tau)\\
& = \displaystyle \sum_{\textbf{d}}C\left(f+\frac{q}{2}\left((\textbf{d}+\textbf{b}')\cdot\textbf{x}+\textbf{d}'\cdot\textbf{x}'
+x_{\gamma}\right),\right. \\ &~~~~~~~~~~~~~~~~~~~~~~~~~~ \left.\tilde{f^*}+\frac{q}{2}\left((\textbf{d}+\textbf{b}'')\cdot\bar{\textbf{x}}+\textbf{d}''\cdot\bar{\textbf{x}'}\right) \right)(\tau)\\
&+\displaystyle \sum_{\textbf{d}}C\left(f+\frac{q}{2}\left((\textbf{d}+\textbf{b}')\cdot\textbf{x}+\textbf{d}'\cdot\textbf{x}'
\right),\right. \\ &~~~~~~~~~~~~~~~~~~~\left. \tilde{f^*}+\frac{q}{2}\left((\textbf{d}+\textbf{b}'')\cdot\bar{\textbf{x}}+\textbf{d}''\cdot\bar{\textbf{x}'}+x_{\gamma}\right) \right)(\tau)\\
&=K_1+K_2,
\end{split}
\end{equation}
\text{where}
\begin{equation}\label{sab4}
\begin{split}
K_1=&\displaystyle\sum_{\textbf{d}}C\left(f+\frac{q}{2}\left((\textbf{d}+\textbf{b}')\cdot\textbf{x}+\textbf{d}'\cdot\textbf{x}'
+x_{\gamma}\right),\right. \\&~~~~~~~~~~~~~~~~~~~ \left. \tilde{f^*}+\frac{q}{2}\left((\textbf{d}+\textbf{b}'')\cdot\bar{\textbf{x}}+
\textbf{d}''\cdot\bar{\textbf{x}'}\right) \right)(\tau),
\end{split}
\end{equation}
and
\begin{equation}\label{sab5}
\begin{split}
K_2=&\displaystyle\sum_{\textbf{d}}C\left(f+\frac{q}{2}\left((\textbf{d}+\textbf{b}')\cdot\textbf{x}+\textbf{d}'\cdot\textbf{x}'
\right),\right. \\& ~~~~~~~~~~~\left. \tilde{f^*}+\frac{q}{2}\left((\textbf{d}+\textbf{b}'')\cdot\bar{\textbf{x}}+
\textbf{d}''\cdot\bar{\textbf{x}'}+x_{\gamma}\right) \right)(\tau).
\end{split}
\end{equation}
The cross-correlation term at RHS of (\ref{sab4}) can be further reduced to
\begin{equation}\nonumber
\begin{split}
&C\left(f+\frac{q}{2}\left((\textbf{d}+\textbf{b}')\cdot\textbf{x}+\textbf{d}'\cdot\textbf{x}'
+x_{\gamma}\right),\right. \\&~~~~~~~~~~~~~~~~~~~~~~~~~~ \left. \tilde{f^*}+\frac{q}{2}\left((\textbf{d}+\textbf{b}'')\cdot\bar{\textbf{x}}+
\textbf{d}''\cdot\bar{\textbf{x}'}\right) \right)(\tau)\\
 &=\displaystyle\sum_{\textbf{c}_1,\textbf{c}_2}C\left(f+\frac{q}{2}\left((\textbf{d}+\textbf{b}')\cdot\textbf{x}+\textbf{d}'\cdot\textbf{x}'
+x_{\gamma}\right)\arrowvert_{\textbf{x}=\textbf{c}_1},\right. \\&~~~~~~~~~~~~~~~~~~~~ \left.\tilde{f^*}+\frac{q}{2}\left((\textbf{d}+\textbf{b}'')\cdot\bar{\textbf{x}}+
\textbf{d}''\cdot\bar{\textbf{x}'}\right)\arrowvert_{\textbf{x}=\textbf{c}_2} \right)(\tau)\qquad\qquad\qquad\qquad\qquad\quad\qquad\qquad\qquad\\
 &=\displaystyle\sum_{\textbf{c}_1,\textbf{c}_2}(-1)^{(\textbf{b}'\cdot\textbf{c}_1)+(\textbf{b}''\cdot\bar{\textbf{c}_2)}}(-1)^{\textbf{d}\cdot(\textbf{c}_1+\bar{\textbf{c}}_2)}
\\&~~~~~~~~~~~~~~~~\times \left(C\left(f+\frac{q}{2}\left(\textbf{d}'\cdot\textbf{x}'
+x_{\gamma}\right)\arrowvert_{\textbf{x}=\textbf{c}_1},\right.\right. \\&~~~~~~~~~~~~~~~~~~~~~~~~~~~~~~~~~\left. \left.\tilde{f^*}+\frac{q}{2}
\textbf{d}''\cdot\bar{\textbf{x}'}\arrowvert_{\textbf{x}=\textbf{c}_2} \right)(\tau)\right).\qquad\qquad\qquad\qquad\qquad\quad\quad\quad\quad
\end{split}
\end{equation}
Therefore (\ref{sab4}) is simplified to
\begin{equation}\label{k_1:1}
\begin{split}
K_1&=\displaystyle\sum_{\textbf{d}}C\left(f+\frac{q}{2}\left((\textbf{d}+\textbf{b}')\cdot\textbf{x}+\textbf{d}'\cdot\textbf{x}'
+x_{\gamma}\right),\right. \\&~~~~~~~~~~~~~~~~~~ \left. \tilde{f^*}+\frac{q}{2}\left((\textbf{d}+\textbf{b}'')\cdot\bar{\textbf{x}}+
\textbf{d}''\cdot\bar{\textbf{x}'}\right) \right)(\tau)\\
&=\displaystyle\sum_{\textbf{d}}
\displaystyle\sum_{\textbf{c}_1,\textbf{c}_2}(-1)^{(\textbf{b}'\cdot\textbf{c}_1)+(\textbf{b}''\cdot\bar{\textbf{c}_2})}(-1)^{\textbf{d}\cdot(\textbf{c}_1+\bar{\textbf{c}}_2)}
\\ &~~~~~~~~~~~~~~\times\left(C\left(f+\frac{q}{2}\left(\textbf{d}'\cdot\textbf{x}'
+x_{\gamma}\right)\arrowvert_{\textbf{x}=\textbf{c}_1},\right.\right. \\ &~~~~~~~~~~~~~~~~~~~~~~~~~~~~~ \left.\left. \tilde{f^*}+\frac{q}{2}
\textbf{d}''\cdot\bar{\textbf{x}'}\arrowvert_{\textbf{x}=\textbf{c}_2} \right)(\tau)\right)\\
&=\displaystyle\sum_{\textbf{c}_1,\textbf{c}_2}(-1)^{(\textbf{b}'\cdot\textbf{c}_1)+(\textbf{b}''\cdot\bar{\textbf{c}_2})}C\left(f\!\!+\!\!\frac{q}{2}\left(\textbf{d}'\cdot\textbf{x}'
\!\!+\!\!x_{\gamma}\right)\arrowvert_{\textbf{x}=\textbf{c}_1},\right. \\&~~~~~~~~~~~~~\left.\tilde{f^*}+\frac{q}{2}
\textbf{d}''\cdot\bar{\textbf{x}'}\arrowvert_{\textbf{x}=\textbf{c}_2} \right)(\tau)\displaystyle\sum_{\textbf{d}}(-1)^{\textbf{d}\cdot(\textbf{c}_1+\bar{\textbf{c}}_2)}.
\end{split}
\end{equation}
By applying \textit{Lemma} \ref{lemma4}, the above can be express as
\begin{equation}\label{sab6}
\begin{split}
K_1&=\displaystyle\mathop{\sum_{\textbf{c}_1,\textbf{c}_2}}_{\textbf{c}_1-\textbf{c}_2=\textbf{1}}2^k(-1)^{(\textbf{b}'\cdot\textbf{c}_1)+(\textbf{b}''\cdot\bar{\textbf{c}_2})}
\qquad\qquad\qquad\qquad\quad\\&~~~~~~~~~~~~~\times\left(C\left(f+\frac{q}{2}\left(\textbf{d}'\cdot\textbf{x}'
+x_{\gamma}\right)\arrowvert_{\textbf{x}=\textbf{c}_1},\right.\right. \\&~~~~~~~~~~~~~~~~~~~~~~~~~~~~~~~\left. \left.\tilde{f^*}+\frac{q}{2}
\textbf{d}''\cdot\bar{\textbf{x}'}\arrowvert_{\textbf{x}=\textbf{c}_2} \right)(\tau)\right).
\end{split}
\end{equation}
The cross-correlation term in (\ref{sab6}) can be simplified to
\begin{equation}
\begin{split}
&C\left(f+\frac{q}{2}\left(\textbf{d}'\cdot\textbf{x}'
+x_{\gamma}\right)\arrowvert_{\textbf{x}=\textbf{c}_1},\tilde{f^*}+\frac{q}{2}
\textbf{d}''\cdot\bar{\textbf{x}'}\arrowvert_{\textbf{x}=\textbf{c}_2} \right)(\tau)\qquad\qquad\qquad\qquad\qquad\qquad\qquad\qquad\qquad\qquad\\
&=\displaystyle\sum_{i=0}^1\sum_{j=0}^1C\left(f+\frac{q}{2}\left(\textbf{d}'\cdot\textbf{x}'
+x_{\gamma}\right)\arrowvert_{\textbf{x}x_{\gamma}=\textbf{c}_1i},\right. \\& ~~~~~~~~~~~~~~~~~~~~~~~~~~~~~~~\left.\tilde{f^*}+\frac{q}{2}
\textbf{d}''\cdot\bar{\textbf{x}'}\arrowvert_{\textbf{x}x_{\gamma}=\textbf{c}_2j} \right)(\tau)\quad\qquad\qquad\qquad\qquad\qquad\qquad\\
&=\displaystyle\sum_{i=0}^1\sum_{j=0}^1(-1)^iC\left(f+\frac{q}{2}\textbf{d}'\cdot\textbf{x}'
\arrowvert_{\textbf{x}x_{\gamma}=\textbf{c}_1i},\right. \\&~~~~~~~~~~~~~~~~~~~~~~~~~~~~~~~ \left.\tilde{f^*}+\frac{q}{2}
\textbf{d}''\cdot\bar{\textbf{x}'}\arrowvert_{\textbf{x}x_{\gamma}=\textbf{c}_2j} \right)(\tau).\quad\qquad\qquad\qquad\qquad\qquad\qquad\\
\end{split}
\end{equation}
Therefore the final expression of (\ref{sab4}) can be expressed  as
\begin{equation}\label{k_1:2}
\begin{split}
K_1 &=\displaystyle\mathop{\sum_{\textbf{c}_1,\textbf{c}_2}}_{\textbf{c}_1-\textbf{c}_2=\textbf{1}}\!\!2^k(-1)^{(\textbf{b}'\cdot\textbf{c}_1)+(\textbf{b}''\cdot\bar{\textbf{c}_2})}
C\left(f\!\!+\!\!\frac{q}{2}\left(\textbf{d}'\cdot\textbf{x}'
\!\!+\!\!x_{\gamma}\right)\arrowvert_{\textbf{x}=\textbf{c}_1},\right. \\ &~~~~~~~~~~~~~~~~~~~~~~~~~~~~~~~~~\left.\tilde{f^*}+\frac{q}{2}
\textbf{d}''\cdot\bar{\textbf{x}'}\arrowvert_{\textbf{x}=\textbf{c}_2} \right)(\tau)\\
 &=\mathop{\sum_{\textbf{c}_1,\textbf{c}_2}}_{\textbf{c}_1-\textbf{c}_2=\textbf{1}}2^k(-1)^{(\textbf{b}'\cdot\textbf{c}_1)+(\textbf{b}''\cdot\bar{\textbf{c}_2})}
\left\{\displaystyle\sum_{i=0}^1\sum_{j=0}^1(-1)^i\right.\\&~~~~~~~~~~~~~~~\times \left. \left(C\left(f+\frac{q}{2}\textbf{d}'\cdot\textbf{x}'
\arrowvert_{\textbf{x}x_{\gamma}=\textbf{c}_1i},\right.\right.\right.\\ &~~~~~~~~~~~~~~~~~~~~~~~~~~~\left.\left.\left.\tilde{f^*}+\frac{q}{2}
\textbf{d}''\cdot\bar{\textbf{x}'}\arrowvert_{\textbf{x}x_{\gamma}=\textbf{c}_2j} \right)(\tau)\right)
\right\}.
\end{split}
\end{equation}
Similarly, (\ref{sab5}) can be simplified to
\begin{equation}\label{k_2:1}
\begin{split}
K_2 &
=\displaystyle\mathop{\sum_{\textbf{c}_1,\textbf{c}_2}}_{\textbf{c}_1-\textbf{c}_2=\textbf{1}}2^k(-1)^{(\textbf{b}'\cdot\textbf{c}_1)+(\textbf{b}''\cdot\bar{\textbf{c}_2})}
C\left(f\!\!+\!\!\frac{q}{2}\textbf{d}'\cdot\textbf{x}'
\arrowvert_{\textbf{x}=\textbf{c}_1},\right. \\&~~~~~~~~~~~~~~~~~~~~~ \left.\tilde{f^*}+\frac{q}{2}\left(
\textbf{d}''\cdot\bar{\textbf{x}'}+x_{\gamma}\right)\arrowvert_{\textbf{x}=\textbf{c}_2} \right)(\tau)\\
 &=\mathop{\sum_{\textbf{c}_1,\textbf{c}_2}}_{\textbf{c}_1-\textbf{c}_2=\textbf{1}}2^k(-1)^{(\textbf{b}'\cdot\textbf{c}_1)+(\textbf{b}''\cdot\bar{\textbf{c}_2})}
\left\{
\displaystyle\sum_{i=0}^1\sum_{j=0}^1(-1)^j\right.\\ &~~~~~~~~~~~~~~~~~~~~~~\times\left.\left(C\left(f+\frac{q}{2}\textbf{d}'\cdot\textbf{x}'
\arrowvert_{\textbf{x}x_{\gamma}=\textbf{c}_1i},\right.\right.\right.\\ &~~~~~~~~~~~~~~~~~~~~~~~~~~\left.\left.\left.\tilde{f^*}+\frac{q}{2}
\textbf{d}''\cdot\bar{\textbf{x}'}\arrowvert_{\textbf{x}x_{\gamma}=\textbf{c}_2j} \right)(\tau)\right)\right\}.
\end{split}
\end{equation}
Therefore, (\ref{sab3}) can be expressed as
\begin{equation}\label{k_2:1}
\begin{split}
&C(\psi(S_{t'}),\psi^*(\bar{S}_{t''}))(\tau)\nonumber\\ &=K_1+K_2\nonumber\\
 &=\mathop{\sum_{\textbf{c}_1,\textbf{c}_2}}_{\textbf{c}_1-\textbf{c}_2=\textbf{1}}2^{k+1}(-1)^{(\textbf{b}'\cdot\textbf{c}_1)+(\textbf{b}''\cdot\bar{\textbf{c}_2})}
\left\{
C\left(f\!\!+\!\!\frac{q}{2}\textbf{d}'\cdot\textbf{x}'
\arrowvert_{\textbf{x}x_{\gamma}=\textbf{c}_10},\right.\right. \\ &~~~~~~~~~~~~~~~~~~~~~~~~~~~~~~~~~~~~\left.\left.\tilde{f^*}+\frac{q}{2}
\textbf{d}''\cdot\bar{\textbf{x}'}\arrowvert_{\textbf{x}x_{\gamma}=\textbf{c}_20} \right)(\tau)\right.\nonumber\\&\left.
~~~~~-C\left(f\!\!+\!\!\frac{q}{2}\textbf{d}'\cdot\textbf{x}'
\arrowvert_{\textbf{x}x_{\gamma}=\textbf{c}_11},\tilde{f^*}\!\!+\!\!\frac{q}{2}
\textbf{d}''\cdot\bar{\textbf{x}'}\arrowvert_{\textbf{x}x_{\gamma}=\textbf{c}_21} \right)(\tau)
\right\}\nonumber\\
 &=\mathop{\sum_{\textbf{c}}}2^{k+1}(-1)^{(\textbf{b}'\cdot\textbf{c}_1)+(\textbf{b}''\cdot\bar{\textbf{c}_2})}
\left\{
C\left(f+\frac{q}{2}\textbf{d}'\cdot\textbf{x}'
\arrowvert_{\textbf{x}x_{\gamma}=\textbf{c}0},\right.\right.\\&~~~~~~~~~~~~~~~~~~~~~~~~~~~~~~~~~~~ \left.\left.\tilde{f^*}+\frac{q}{2}
\textbf{d}''\cdot\bar{\textbf{x}'}\arrowvert_{\textbf{x}x_{\gamma}=\textbf{c+1}0} \right)(\tau)\right.\quad\nonumber\\ &\left.
~~~-C\left(f+\frac{q}{2}\textbf{d}'\cdot\textbf{x}'
\arrowvert_{\textbf{x}x_{\gamma}=\textbf{c}1},\tilde{f^*}+\frac{q}{2}
\textbf{d}''\cdot\bar{\textbf{x}'}\arrowvert_{\textbf{x}x_{\gamma}=\textbf{c+1}1} \right)(\tau)
\right\}\\
&=\mathop{\sum_{\textbf{c}}}2^{k+1}(-1)^{(\textbf{b}'\cdot\textbf{c}_1)+(\textbf{b}''\cdot\bar{\textbf{c}_2})}\\ &~~~~~~~~~~~~~~~~~~\times
\left\{
\mathop{\sum_{\textbf{c}_1',\textbf{c}_2'}}\!\!\left(C\left(f+\frac{q}{2}\textbf{d}'\cdot\textbf{x}'
\arrowvert_{\textbf{x}x_{\gamma}\textbf{x}'=\textbf{c}0\textbf{c}_1'},\right.\right.\right.~~~~~\\&~~~~~~~~~~~~~~~~~~~~~~~~~~~ \left.\left. \left.\tilde{f^*}+\frac{q}{2}
\textbf{d}''\cdot\bar{\textbf{x}'}\arrowvert_{\textbf{x}x_{\gamma}\textbf{x}'=\textbf{c+1}0\textbf{c}_2'} \right)(\tau)\quad\right.\right.\\ &\left.\left.
~~~-C\left(f+\frac{q}{2}\textbf{d}'\cdot\textbf{x}'
\arrowvert_{\textbf{x}x_{\gamma}\textbf{x}'=\textbf{c}1\textbf{c}_1'},\right.\right.\right.\\&~~~~~~~~~~~~~~~~~~~~~\left.\left. \left. \tilde{f^*}+\frac{q}{2}
\textbf{d}''\cdot\bar{\textbf{x}'}\arrowvert_{\textbf{x}x_{\gamma}\textbf{x}'=\textbf{c+1}1\textbf{c}_2'} \right)(\tau)\right)
\right\}\\
&=\mathop{\sum_{\textbf{c}}}2^{k+1}(-1)^{(\textbf{b}'\cdot\textbf{c}_1)+(\textbf{b}''\cdot\bar{\textbf{c}_2})}
\left\{
\mathop{\sum_{\textbf{c}_1',\textbf{c}_2'}}(-1)^{(\textbf{d}'\cdot\textbf{c}_1')+(\textbf{d}''\cdot\bar{\textbf{c}}_2'')}\right. \\ &~~~~~~~~~~~~\times\left.
\left(C\left(f\arrowvert_{\textbf{x}x_{\gamma}\textbf{x}'=\textbf{c}0\textbf{c}_1'},\tilde{f^*}\arrowvert_{\textbf{x}x_{\gamma}\textbf{x}'=\textbf{c+1}0\textbf{c}_2'} \right)(\tau)\right.\right.\\ &\left.\left.
~~~~~~~~~~~~~~~~~~-C\left(f\arrowvert_{\textbf{x}x_{\gamma}\textbf{x}'=\textbf{c}1\textbf{c}_1'},\tilde{f^*}\arrowvert_{\textbf{x}x_{\gamma}\textbf{x}'=\textbf{c+1}1\textbf{c}_2'} \right)(\tau)\right)
\right\}.
\end{split}
\end{equation}
Since $G(f\arrowvert_{\textbf{x}=\textbf{c}})$ contains a path over $m-k-p$ vertices and $p$ isolated vertices, the Boolean function $f\arrowvert_{\textbf{x}=\textbf{c}}$
can be expressed as
\begin{equation}\nonumber
\begin{split}
f\arrowvert_{\textbf{x}=\textbf{c}}=&\frac{q}{2}\displaystyle\sum_{\alpha=0}^{m-k-p-2}x_{l_{\alpha}}x_{l_{\alpha+1}}+\displaystyle\sum_{\alpha=0}^{m-k-p-1}g_{l_{\alpha}}x_{l_{\alpha}}\\&+\displaystyle\sum_{j=1}^{p}g_{m-p-1+j}x_{m-p-1+j}+g'.
\end{split}
\end{equation}
Let $h_1$ denotes the function obtained from $f$ by substituting $\textbf{x}=\textbf{c}$,
 $\textbf{x}'=\textbf{c}'_1$ and $x_{\gamma}=1$ for some binary vectors $\textbf{c}$ and $\textbf{c}'_1$ and let $h_2$ be the corresponding function when
 $\textbf{x}=\textbf{c}$,  $\textbf{x}'=\textbf{c}_1'$ and $x_{\gamma}=0$. Further we assume that $\gamma=l_{m-k-p-1}$ without loss of generality.
   Then the function $h_1$ and $h_2$ can be expressed as\\
\begin{equation}\nonumber
\begin{split}
h_1=&\frac{q}{2}\displaystyle\sum_{\alpha=0}^{m-k-p-3}x_{l_{\alpha}}x_{l_{\alpha+1}}+\displaystyle\sum_{\alpha=0}^{m-k-p-2}g_{l_{\alpha}}x_{l_{\alpha}}+
\varGamma\cdot\textbf{c}'_1\\&+\frac{q}{2}x_{l_{m-k-p-2}}+g_{l_{m-k-p-1}}+g',\\
h_2=&\frac{q}{2}\displaystyle\sum_{\alpha=0}^{m-k-p-3}x_{l_{\alpha}}x_{l_{\alpha+1}}+\displaystyle\sum_{\alpha=0}^{m-k-p-2}g_{l_{\alpha}}x_{l_{\alpha}}+
\varGamma\cdot\textbf{c}'_1+g'.\qquad\qquad\qquad\qquad\qquad
\end{split}
\end{equation}
Similarly, the nonzero components of the complex vectors $\textbf{a}=\psi(f\arrowvert_{\textbf{x}x_{\gamma}\textbf{x}'=\textbf{c}1\textbf{c}_1'})$ and
$\textbf{b}=\psi(f\arrowvert_{\textbf{x}x_{\gamma}\textbf{x}'=\textbf{c}0\textbf{c}_1'})$ are given by the functions $h_1$ and $h_2$ respectively.
Let $\textbf{c}$ and $\textbf{d}$ be two complex vectors whose nonzero components are obtained from the
functions $h_1-\varGamma\cdot\textbf{c}_1'$ and $h_2-\varGamma\cdot\textbf{c}_1'$. Therefore, $\textbf{a}=\omega^{\varGamma\cdot\textbf{c}_1'}\textbf{c}$
and $\textbf{b}=\omega^{\varGamma\cdot\textbf{c}_1'}\textbf{d}$.\\
Similarly, the nonzero components of the vectors $\textbf{a}_1=\psi^*(\tilde{f}\arrowvert_{\textbf{x}x_{\gamma}\textbf{x}'=\textbf{c+1}0\textbf{c}_2'})$
and $\textbf{b}_1=\psi^*(\tilde{f}\arrowvert_{\textbf{x}x_{\gamma}\textbf{x}'=\textbf{c+1}1\textbf{c}_2'})$ are obtained
by the functions
\begin{equation}\nonumber
\begin{split}
h_3=&\frac{q}{2}\displaystyle\sum_{\alpha=0}^{m-k-p-3}(1\!-\!x_{l_{\alpha}})(1\!-\!x_{l_{\alpha+1}})\!+\!\!\displaystyle\sum_{\alpha=0}^{m-k-p-2}g_{l_{\alpha}}(1-x_{l_{\alpha}})\\&+
\varGamma\cdot\bar{\textbf{c}'_2}+\frac{q}{2}(1-x_{l_{m-k-p-2}})+g_{l_{m-k-p-1}}+g'\\
=&\tilde{h}_1-\varGamma\cdot(\bar{\textbf{c}}_1')+\varGamma\cdot\bar{\textbf{c}'_2},\qquad\qquad\qquad\qquad\qquad\qquad\qquad
\qquad\qquad\qquad\qquad\qquad\qquad
\end{split}
\end{equation}
\begin{equation}\nonumber
\begin{split}
h_4=&\frac{q}{2}\displaystyle\sum_{\alpha=0}^{m-k-p-3}(1\!-\!x_{l_{\alpha}})(1\!-\!x_{l_{\alpha+1}})\!+\!\!\displaystyle\sum_{\alpha=0}^{m-k-p-2}g_{l_{\alpha}}(1\!-\!x_{l_{\alpha}})\\
&+
\varGamma\cdot\bar{\textbf{c}'}_2+g'\qquad\qquad\qquad\qquad\qquad\qquad\qquad\quad\\
=&\tilde{h}_2-\varGamma\cdot(\bar{\textbf{c}}_1')+\varGamma\cdot\bar{\textbf{c}'}_2.\qquad\qquad\qquad\qquad\qquad\qquad
\qquad\qquad\qquad\qquad\qquad\qquad\qquad\qquad\qquad\qquad\qquad\quad
\end{split}
\end{equation}
Therefore $\textbf{a}_1=\omega^{\varGamma\cdot\bar{c}'_2}\tilde{\textbf{c}}^*$ and $\textbf{b}_1=\omega^{\varGamma\cdot\bar{c}'_2}\tilde{\textbf{d}}^*$.\\
Now, the difference of cross-correlation terms of (\ref{k_2:1}) can be simplified to
\begin{equation}\label{con}
\begin{split}
C&\left(f\arrowvert_{\textbf{x}x_{\gamma}\textbf{x}'=\textbf{c}0\textbf{c}_1'},\tilde{f^*}\arrowvert_{\textbf{x}x_{\gamma}\textbf{x}'=\textbf{c+1}0\textbf{c}_2'} \right)(\tau)\\
&~~~~~~~~~~~~~-C\left(f\arrowvert_{\textbf{x}x_{\gamma}\textbf{x}'=\textbf{c}1\textbf{c}_1'},\tilde{f^*}\arrowvert_{\textbf{x}x_{\gamma}\textbf{x}'=\textbf{c+1}1\textbf{c}_2'} \right)(\tau)\\
=&C(\textbf{b},\textbf{a}_1)(\tau)-C(\textbf{a},\textbf{b}_1)(\tau)\qquad\qquad\qquad\qquad\qquad\qquad\qquad\qquad\qquad\quad\\
=&\omega^{\varGamma\cdot\textbf{c}_1'}\bar{\omega}^{\varGamma\cdot\bar{c}'_2}\left(C(\textbf{d},\tilde{\textbf{c}}^*)(\tau)-C(\textbf{c},\tilde{\textbf{d}}^*)(\tau)\right).
\qquad\qquad\qquad\qquad\qquad\qquad\quad
\end{split}
\end{equation}
For any two complex sequences $\textbf{c}$ and $\textbf{d}$, recall the identity
\begin{equation}\nonumber
C(\textbf{c},\tilde{\textbf{d}}^*)(\tau)=C(\tilde{\textbf{d}},\textbf{c}^*)(-\tau)=C(\textbf{d},\tilde{\textbf{c}}^*)(\tau).
\end{equation}
Therefore, substituting (\ref{con}) in (\ref{k_2:1}) and  using the above identity we have $K_1+K_2=0$, thus completing the proof.

\end{IEEEproof}

\section{Proof of \textit{Remark 1}}
\begin{IEEEproof}
In Fig. 1, if we fix the set $X_S=\{x_{i_1},x_{i_2},$ $\cdots,x_{i_p}\}$ in
$\{x_{m-p},x_{m-p+1},$ $\cdots,x_{m-1}\}$, then a GBF corresponding to Fig. 1 produces an optimal ZCCS. Our task is to find out the number of such
distinct GBFs. After fixing $X_S=\{x_{m-p},x_{m-p+1},$ $\cdots,x_{m-1}\}$, the set $X_J=\{x_{j_0},x_{j_1},\cdots,x_{j_{k-1}}\}$ can be
chosen in $m-p\choose k$ ways. For each choice of $X_J$, the set $X_P$ can be chosen in only one way. The quadratic form $Q$ given in (17), can
be expressed as $Q=Q_1+Q_2+Q_3+Q_4$ where
\begin{equation}
\begin{split}
Q_1&=\displaystyle\frac{q}{2}\sum_{i=0}^{m-k-p-2}x_{l_i}x_{l_{i+1}},\\
Q_2&=\displaystyle\sum_{i=0}^{m-k-p-1}\sum_{\alpha=0}^{k-1}a'_{i,\alpha}x_{l_i}x_{j_\alpha},\\
Q_3&=\displaystyle\sum_{\alpha=0}^{k-1}\sum_{\beta=1}^{p}e'_{\alpha,\beta}x_{j_\alpha}x_{i_\beta},\\
Q_4&=\displaystyle\sum_{0\leq \alpha_1<\alpha_2<k}b'_{\alpha_1,\alpha_2}x_{j_{\alpha_1}}x_{j_{\alpha_2}}.
\end{split}
\end{equation}
For each choice of $m-p\choose k$, we get $\frac{(m-k-p)!}{2}$ distinct $Q_1$, $(q-1)^{k(m-k-p)}$ distinct $Q_2$, $q^{kp}$ distinct $Q_3$, and
$q^{\frac{k(k-1)}{2}}$ distinct $Q_4$. Finally, we get at least
\begin{equation}
\frac{(m-p)!}{2(k!)}(q-1)^{k(m-k-p)}q^{kp+\frac{k(k-1)}{2}}
\end{equation}
distinct quadratic forms. Corresponding to each quadratic form $Q$, we get $q^{m+1}$ distinct GBFs. Therefore there exist at least $\frac{(m-p)!}{2(k!)}(q-1)^{k(m-k-p)}q^{kp+\frac{k(k-1)}{2}+m+1}$
distinct GBFs corresponding to which we get the same number of distinct optimal ZCCSs. In the above enumeration, we have taken $a'_{i,\alpha}\in \mathbb{Z}_q\setminus\{\frac{q}{2}\}$, otherwise
sometimes we can get some ZCCSs more than once.
\end{IEEEproof}
\end{appendices}
\bibliographystyle{IEEEtran}
\bibliography{notesk}

\begin{thebibliography}{10}
\providecommand{\url}[1]{#1}
\csname url@samestyle\endcsname
\providecommand{\newblock}{\relax}
\providecommand{\bibinfo}[2]{#2}
\providecommand{\BIBentrySTDinterwordspacing}{\spaceskip=0pt\relax}
\providecommand{\BIBentryALTinterwordstretchfactor}{4}
\providecommand{\BIBentryALTinterwordspacing}{\spaceskip=\fontdimen2\font plus
\BIBentryALTinterwordstretchfactor\fontdimen3\font minus
  \fontdimen4\font\relax}
\providecommand{\BIBforeignlanguage}[2]{{%
\expandafter\ifx\csname l@#1\endcsname\relax
\typeout{** WARNING: IEEEtran.bst: No hyphenation pattern has been}%
\typeout{** loaded for the language `#1'. Using the pattern for}%
\typeout{** the default language instead.}%
\else
\language=\csname l@#1\endcsname
\fi
#2}}
\providecommand{\BIBdecl}{\relax}
\BIBdecl

\bibitem{chen2007next}
H.-H. Chen, \emph{The Next Generation CDMA Technologies}.\hskip 1em plus 0.5em
  minus 0.4em\relax Wiley, 2007.

\bibitem{gol1961}
M.~Golay, ``Complementary series,'' \emph{IRE Trans. Inf. Theory}, vol.~7,
  no.~2, pp. 82--87, Apr. 1961.

\bibitem{chong1972}
C.-C. Tseng and C.~Liu, ``Complementary sets of sequences,'' \emph{IEEE Trans.
  Inf. Theory}, vol.~18, no.~5, pp. 644--652, Sep. 1972.

\bibitem{Davis1999}
J.~A. Davis and J.~Jedwab, ``Peak-to-mean power control in {OFDM}, {Golay}
  complementary sequences, and {Reed-Muller} codes,'' \emph{IEEE Trans. Inf.
  Theory}, vol.~45, no.~7, pp. 2397--2417, Nov. 1999.

\bibitem{Li2010}
Y.~Li, ``A construction of general {QAM} {Golay} complementary sequences,''
  \emph{IEEE Trans. Inf. Theory}, vol.~56, no.~11, pp. 5765--5771, Nov. 2010.

\bibitem{Liu2013}
Z.~Liu, Y.~Li, and Y.~L. Guan, ``New constructions of general {QAM} {Golay}
  complementary sequences,'' \emph{IEEE Trans. Inf. Theory}, vol.~59, no.~11,
  pp. 7684--7692, Nov. 2013.

\bibitem{pater2000}
K.~G. Paterson, ``Generalized {Reed-Muller} codes and power control in {OFDM}
  modulation,'' \emph{IEEE Trans. Inf. Theory}, vol.~46, no.~1, pp. 104--120,
  Jan. 2000.

\bibitem{kusch}
K.~U. Schmidt, ``Complementary sets, generalized {Reed-Muller} codes, and power
  control for {OFDM},'' \emph{IEEE Trans. Inf. Theory}, vol.~53, no.~2, pp.
  808--814, Feb. 2007.

\bibitem{arthina}
A.~Rathinakumar and A.~K. Chaturvedi, ``Complete mutually orthogonal {G}olay
  complementary sets from {Reed-Muller} codes,'' \emph{IEEE Trans. Inf.
  Theory}, vol.~54, no.~3, pp. 1339--1346, Mar. 2008.

\bibitem{uda2014}
Z.~Liu, Y.~L. Guan, and U.~Parampalli, ``New complete complementary codes for
  peak-to-mean power control in multi-carrier {CDMA},'' \emph{IEEE Trans.
  Commun.}, vol.~62, no.~3, pp. 1105--1113, Mar. 2014.

\bibitem{Liu_FDRR_2015}
Z.~Liu, Y.~L. Guan, and H.-H. Chen, ``Fractional-delay-resilient receiver
  design for interference-free {MC-CDMA} communications based on complete
  complementary codes,'' \emph{IEEE Trans. Wireless Commun.}, vol.~14, no.~3,
  pp. 1226--1236, Mar. 2015.

\bibitem{cccsmajhi}
S.~Das, S.~Budi\v{s}in, S.~Majhi, Z.~Liu, and Y.~L. Guan, ``A multiplier-free
  generator for polyphase complete complementary codes,'' \emph{IEEE Trans.
  Signal Process.}, vol.~66, no.~5, pp. 1184--1196, Mar. 2018.

\bibitem{slett}
S.~Das, S.~Majhi, and Z.~Liu, ``A novel class of complete complementary codes
  and their applications for apu matrices,'' \emph{IEEE Signal Process. Lett.},
  vol.~25, no.~9, pp. 1300--1304, Sept. 2018.

\bibitem{liu2011}
Z.~Liu, Y.~L. Guan, B.~C. Ng, and H.-H. Chen, ``Correlation and set size bounds
  of complementary sequences with low correlation zone,'' \emph{IEEE Trans.
  Commun.}, vol.~59, no.~12, pp. 3285--3289, Dec. 2011.

\bibitem{Liu-ITW2014}
Z.~Liu, Y.~L. Guan, and U.~Parampalli, ``A new construction of zero correlation
  zone sequences from generalized {Reed-Muller} codes,'' in \emph{2014 IEEE
  Information Theory Workshop}, Nov. 2014, pp. 591--595.

\bibitem{fan2007}
P.~Fan, W.~Yuan, and Y.~Tu, ``Z-complementary binary sequences,'' \emph{IEEE
  Signal Process. Lett.}, vol.~14, no.~8, pp. 509--512, Aug. 2007.

\bibitem{lfengispl2008}
L.~Feng, P.~Fan, X.~Tang, and K.~K. Loo, ``Generalized pairwise
  {Z}-complementary codes,'' \emph{IEEE Signal Process. Lett.}, vol.~15, pp.
  377--380, 2008.

\bibitem{jli_igc_2008}
J.~Li, A.~Huang, M.~Guizani, and H.-H. Chen, ``Inter group complementary codes
  for interference resistant {CDMA} wireless communications,'' \emph{IEEE
  Trans. Wireless Commun.}, vol.~7, no.~1, pp. 166--174, Jan. 2008.

\bibitem{sarkar_igc}
P.~Sarkar, S.~Majhi, H.~Vettikalladi, and A.~S. Mahajumi, ``A direct
  construction of inter-group complementary code set,'' \emph{IEEE Access}, pp.
  1--1, 2018.

\bibitem{fan2008}
W.~Yuan, Y.~Tu, and P.~Fan, ``Optimal training sequences for
  cyclic-prefix-based single-carrier multi-antenna systems with space-time
  block-coding,'' \emph{IEEE Trans. Wireless Commun.}, vol.~7, no.~11, pp.
  4047--4050, Nov. 2008.

\bibitem{hmwang2007}
H.~M. Wang, X.~Q. Gao, B.~Jiang, X.~H. You, and W.~Hong, ``Efficient {MIMO}
  channel estimation using complementary sequences,'' \emph{IET Commun.},
  vol.~1, no.~5, pp. 962--969, Oct. 2007.

\bibitem{16qamliu}
Z.~Liu and Y.~L. Guan, ``16-{QAM} almost-complementary sequences with low
  {PMEPR},'' \emph{IEEE Trans. Commun.}, vol.~64, no.~2, pp. 668--679, Feb.
  2016.

\bibitem{stinch}
T.~E. Stinchcombe, ``Aperiodic correlations of length $2^m$ sequences,
  complementarity, and power control for {OFDM},'' Ph.D. dissertation,
  University of London, Apr. 2000.

\end{thebibliography}
\end{document}